\documentclass[usenatbib,usegraphicx]{mn2e}
\usepackage{times}
\usepackage{graphicx, epsfig}
\usepackage[fleqn]{amsmath}
\usepackage{amsfonts}

\title[Dynamical modelling of NGC~2974]{Dynamical modelling of stars and gas 
in NGC~2974:\\ determination of mass-to-light ratio, inclination and\\ 
orbital structure by Schwarzschild's method}
\author[Davor Krajnovi\'c et al.]
       {Davor Krajnovi\'c,$^1$\thanks{E-mail: davor@strw.leidenuniv.nl}
        Michele Cappellari,$^1$
        Eric Emsellem,$^2$\newauthor
		Richard M. McDermid,$^1$
        P. Tim de Zeeuw$^1$\\
$^1$Sterrewacht Leiden, Postbus 9513, 2300 RA Leiden, The Netherlands\\
$^2$Centre de Recherche Astronomique de Lyon, 9 Avenue Charles
    Andr\'e, 69230 Saint-Genis-Laval, France}

\pagerange{\pageref{firstpage}--\pageref{lastpage}}
\pubyear{2004}

\newcommand{\SAURON}{{\tt SAURON} }
\newcommand{\kms}{\>{\rm km}\,{\rm s}^{-1}}
\newcommand{\Msun}{\>{\rm M_{\odot}}}
\newcommand{\ud}{\mathrm{d}}

\def\aj{AJ}             
\def\apj{ApJ}           
\def\apjl{ApJ}          
\def\apjs{ApJS}         
\def\aap{A\&A}          
\def\aaps{A\&AS}        
\def\mnras{MNRAS}       
\def\pasp{PASP}         

\begin{document}
\label{firstpage}
\maketitle

\begin{abstract}
We study the large-scale stellar and gaseous kinematics of the E4
galaxy NGC~2974, based on panoramic integral-field data obtained with
\SAURON. We quantify the velocity fields with Fourier methods
(kinemetry), and show that the large-scale kinematics is largely
consistent with axisymmetry. We construct general axisymmetric
dynamical models for the stellar motions using Schwarzschild's
orbit-superposition method, and compare the inferred inclination and
mass-to-light ratio with the values obtained by modelling the gas
kinematics. Both approaches give consistent results. However we find
that the stellar models provide fairly weak constraints on the
inclination. The intrinsic orbital distribution of NGC~2974, which we
infer from our model, is characterised by a large-scale stellar
component of high angular momentum.  We create semi-analytic test
models, resembling NGC~2974, to study the ability of Schwarzschild's
modelling technique to recover the given input parameters
(mass-to-light ratio and inclination) and the distribution
function. We also test the influence of a limited spatial coverage on
the recovery of the distribution function (i.e. the orbital
structure). We find that the models can accurately recover the input
mass-to-light ratio, but we confirm that even with perfect input
kinematics the inclination is only marginally constrained. This
suggests a possible degeneracy in the determination of the
inclination, but further investigations are needed to clarify this
issue. For a given potential, we find that the analytic distribution
function of our test model is well recovered by the three-integral
model within the spatial region constrained by integral-field
kinematics.
\end{abstract}


%
%

\section{Introduction}
\label{s:intro}
The internal dynamical structure of galaxies retains evidence of their
evolution. The internal dynamics, however, can only be interpreted
through a combination of observational and theoretical efforts. From a
theoretical point of view, one wants to know how the stars are
distributed in space and what velocities they have. From the
observational point of view, one wants to determine the intrinsic
structure of the observed galaxies. The goals of both approaches are
equivalent, and consist of the recovery of the phase-space density, or
distribution function (DF) of galaxies, which uniquely specifies their
properties. An insight into the DF is possible by the construction of
dynamical models which are constrained by observations. There are
several modelling methods established in the literature, of which
Schwarzschild's orbit-superposition method is perhaps the most elegant
\citep{1979ApJ...232..236S, 1982ApJ...263..599S}. In the past few
years it has been applied successfully to a number of galaxies
\citep{1998ApJ...493..613V, 1999ApJ...514..704C, 2002ApJ...578..787C,
2003ApJ...583...92G}; but recent observational advances in
spectroscopy with integral-field units offer for the first time full
two-dimensional constraints on these dynamical models
(\citealt{2002MNRAS.335..517V}; Copin, Cretton \& Emsellem
2004\nocite{2004A&A...415..889C}).

This paper presents a case study of the early-type galaxy NGC~2974. It
is one of the few elliptical galaxies known to contain an extended
disc of neutral hydrogen in regular rotation
\citep{1988ApJ...330..684K}. It also hosts extended H$\alpha$ emission
\citep{1993A&A...280..409B, 1998A&AS..128...75P}, and belongs to the
`rapid rotators' \citep{1988A&A...193L...7B}. The total absolute
magnitude of $M_B=-20.32$ puts NGC~2974 near the transition between
giant ellipticals and the lower-luminosity objects which often show
photometric and kinematic evidence for a significant disc component
\citep[e.g.][]{1992MNRAS.254..389R}.  \citet[][hereafter
EGF03]{2003MNRAS.345.1297E}, combining WFPC2 imaging with TIGER
integral-field spectroscopy of the central few arcseconds, discovered
a spiral structure in the H$\alpha$ emission in the inner few
arcseconds, and concluded that the galaxy contains a inner stellar
bar. The general properties of NGC~2974 are listed in
Table~\ref{t:prop}.

The availability of both stellar and gaseous kinematics makes NGC~2974
a very interesting case for detailed dynamical
modelling. \citet{1994MNRAS.270..325C} made dynamical Jeans models of
the gaseous and stellar components additionally introducing a stellar
disc in order to fit their long-slit data along three position
angles. They found that the stellar and gaseous discs were
kinematically aligned and the inclination of both discs was consistent
with 60\degr. This prompted them to suggest a common evolution, where
the gas could be ionised by the stars in the stellar disc. Using more
sophisticated two-integral axisymmetric models, which assume the DF
depends only on the two classical integrals of motion, the energy $E$
and the angular momentum with respect to the symmetry axis $L_z$,
EGF03\nocite{2003MNRAS.345.1297E} were able to reproduce all features
of \citet{1994MNRAS.270..325C} data as well as their integral-field
{\tt TIGER} data (covering the inner 4\arcsec). The models of
EGF03\nocite{2003MNRAS.345.1297E} did not require a thin stellar disc
to fit the data.

In this study we construct axisymmetric models for NGC~2974 based on
Schwarzschild's orbit superposition method. This method allows the DF
to depend on all three isolating integrals of motion. All previous
studies with three-integral models concentrated on the determination
of the mass-to-light ratio $\Upsilon$, and mass of the central black
hole, $M_{\rm BH}$. Based on the observed stellar velocity dispersion,
the $M_{\rm BH} - \sigma$ relation \citep[e.g.][]{2002ApJ...574..740T}
predicts a central black hole mass of $2.5 \times 10^8 M_\odot$, which
at the distance of NGC~2974 (21.48 Mpc, \citealt{2001ApJ...546..681T})
has a radius of influence of $0\farcs2$. Our observations of NGC~2974,
with the integral-field spectrograph \SAURON
\citep{2001MNRAS.326...23B}, do not have the necessary resolution to
probe the sphere of influence of the central black hole. The dynamical
models presented here are therefore aimed at determination of the
$\Upsilon$, the inclination, $i$, and the internal orbital
structure. The stellar and gaseous kinematics also provide independent
estimates of $\Upsilon$ and $i$, which can be used to cross-validate
the results from the two approaches.

The results of the dynamical modelling are influenced by the
assumptions of the models, but also by the specifics of the
observations. The spatial coverage of the kinematics is one
example. The two-dimensional coverage is an improvement over a few
slits often used in other studies. Similarly, increasing the radial
extent of the data could change the results.  Another issue,
associated with the modelling techniques, is the ability of the
three-integral models to recover the true distribution function of the
galaxy. This is very important for the investigation of the internal
dynamics, since the recovered orbital distribution must represent the
observed galaxy if we want to learn about the galaxy's evolutionary
history. In this paper we present tests designed to probe these issues
and, in general, to determine the robustness of our three-integral
method.

This paper is organised as follows. Section~\ref{s:obs} summarises the
\SAURON spectroscopy and the photometric ground- and space-based
data. The analysis of the velocity maps, used to quantify the presence
and influence of possible non-axisymmetric motions as well as a brief
discussion on bars in NGC~2974, is presented in
Section~\ref{s:kin}. The three-integral dynamical models for the
stellar motions are discussed in
Section~\ref{s:dyno}. Section~\ref{s:tests} is devoted to tests of the
three-integral method involving the determination of the model
parameters ($\Upsilon$, $i$), influence of the radial extent of the
data and the recovery of the DF. The modelling of the emission-line
gas kinematics and comparison with the results of the stellar
dynamical modelling is presented in Section~\ref{s:em_gas}.
Section~\ref{s:con} concludes.

%
%

\section{Observations and data reduction}
\label{s:obs}
The observations of NGC~2974 used in this work consist of ground- and
space-based imaging, and ground-based integral-field spectroscopy. The
imaging data were presented in EGF03\nocite{2003MNRAS.345.1297E} and
the absorption-line kinematics of the \SAURON observations in
\citet[][hereafter E04]{2004MNRAS.352..721E} as part of the \SAURON
survey \citep{2002MNRAS.329..513D}. In this study we also use the
\SAURON emission-line kinematics of NGC~2974.

\subsection{\SAURON spectroscopy}
\label{ss:spec}
NGC~2974 was observed with the integral-field spectrograph \SAURON
mounted on the 4.2-m William Herschel Telescope (WHT) in March
2001. The observations consisted of eight exposures divided equally
between two pointings, each covering the centre and one side
of the galaxy. The individual exposures of both pointings were
dithered to obtain a better estimate of detector sensitivity
variations and avoid systematic errors. The instrumental
characteristics of \SAURON and a summary of the observations are
presented in Table~\ref{t:sau_ins}.

The \SAURON~data were reduced following the steps described in
\citet{2001MNRAS.326...23B} using the dedicated software {\tt XSauron}
developed at CRAL-Observatoire.  The performed reduction
steps included bias and dark subtraction, extraction of the spectra
using a fitted mask model, wavelength calibration, low frequency
flat-fielding, cosmic-ray removal, homogenisation of the spectral
resolution over the field, sky subtraction and flux calibration. All
eight exposures were merged into one data cube with a common
wavelength range by combining the science and noise spectra using
optimal weights and (re)normalisation. In this process we resampled
the data-cube to a common spatial scale ($0\farcs8\times0\farcs8$)
with a resulting field of view of about $45\arcsec \times
45\arcsec$. The data cube was spatially binned to increase the
signal-to-noise ($S/N$) ratio over the field, using the Voronoi 2D
binning algorithm of \citet{2003MNRAS.342..345C}. The targeted minimum
$S/N$ was 60 per aperture, but most of the spectra have $S/N$ ratio
high (e.g. [$S/N$]$_{max}\approx 420$) and about half of the spatial
elements remain un-binned. The final data cube of NGC~2974 and the
detailed reduction procedure was presented in E04.

\begin{table}
 \caption[]{Properties of NGC~2974.}
  \label{t:prop}
$$
  \begin{array}{lr}
    \hline
    $Parameter$ & $Value$\\
    \hline
    \noalign{\smallskip}
    $Morphological type$                 & $E4$\\
    $M$_{B}$ [mag]$                      & -20.46\\
    $effective B-V [mag]$                & 1.00\\
    $PA [degrees]$                       & 42\\
    $Distance Modulus [mag]$             & 31.66\\
    $Distance scale [pc/arcsec]$         & 104.13 \\
    \noalign{\smallskip}
    \hline
  \end{array}
$$
{Notes -- Listed properties are taken from the Lyon/Meudon
Extragalactic Database (LEDA). Distance modulus is from
\citet{2001ApJ...546..681T}}
\end{table}

\begin{figure*}
        \includegraphics[width=\textwidth]{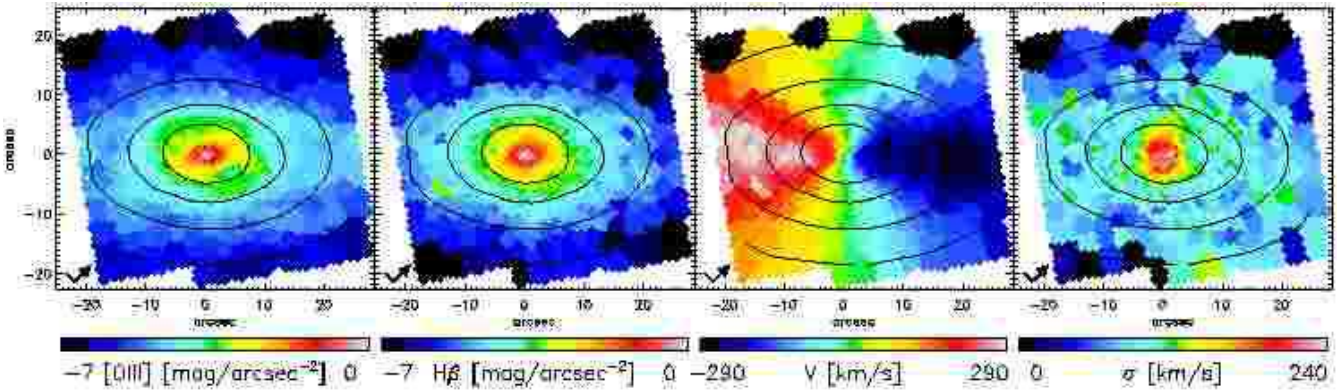}
        \caption{\label{f:gas_maps} Distribution of H$\beta$ and
        [O{\small III}] emission lines and gas kinematics observed by
        {\tt SAURON}. Gas intensities are in mag arcsec$^{-2}$ with
        arbitrary zero points. Gas mean velocity V and velocity
        dispersion $\sigma$ are in $\kms$. Overplotted contours are
        levels of the reconstructed total intensity from the full
        \SAURON spectra. Arrow in the left corner of each plot marks
        North-East orientation of the maps. }
\end{figure*}

\begin{table}
 \caption[]{The \SAURON instrumental characteristics and exposure
 details of the observations of NGC~2974 obtained in March 2001
 at the WHT. The sky apertures are pointed 1\farcm9 away from the main
 field.}
  \label{t:sau_ins}
$$
  \begin{array}{ll}
    \hline
    \noalign{\smallskip}
    $Field of view$                      & 33\arcsec \times 41\arcsec\\
    $Aperture size$                      & 0\farcs94\\
    $Final spatial sampling$             & 0\farcs8\\
    $Spectral range$                     & 4810 - 5300 $\AA$\\
    $Spectral sampling$                  & 1.1 $\AA\, pixel$^{-1}\\
    $Spectral resolution$                & 4.2 $\AA\, (FWHM)$ \\
    $\# of field lenses$                 & 1431\\
    $\# of sky lenses$                   & 146\\
    $\# of exposures$                    & 8\\
    $Exposure time per pointing$         & 1800\, $s$\\
    $Instrumental dispersion$\, (\sigma) & 108 \kms\\
    $Median seeing (FWHM)$               &1\farcs4\\
    \noalign{\smallskip}
    \hline
  \end{array}
$$
\end{table}

\subsection{Absorption-line kinematics}
\label{ss:kin}
The \SAURON spectral range includes several important emission lines:
H$\beta$, [O{\small III}]$\lambda\lambda$4959,5007 and
[NI]$\lambda\lambda$5198,5200 doublets. These lines have to be masked
or removed from the spectra used for the extraction of the stellar
kinematics. The method most suitable for this is the direct
pixel-fitting method operating in wavelength space, which allows easy
masking of the emission lines. We used the penalised pixel-fitting
algorithm (pPXF) of \citet{2004PASP..116..138C}, following the
prescriptions of E04. The line-of-sight velocity distribution (LOSVD)
was parametrised by the Gauss-Hermite expansion
\citep{1993ApJ...407..525V, 1993MNRAS.265..213G}. The 2D stellar
kinematic maps of NGC~2974, showing the mean velocity (V), the
velocity dispersion ($\sigma$), as well as higher order Gauss-Hermite
moments $h_{3}$ and $h_{4}$, were presented in
E04\nocite{2004MNRAS.352..721E}, along with the kinematics of 47 other
elliptical and lenticular galaxies. In this study we expand on the
previously published kinematics by including two more terms in the
Gauss-Hermite expansion ($h_{5}$ and $h_{6}$) to make sure all useful
information was extracted from the spectra and tighten the constraints
on the dynamical models. The extraction of additional kinematic terms
was performed following the same procedure as in E04. The new
extraction is consistent with the published kinematics (except that
now the LOSVD is parameterised with 6 moments) and we do not present
them here explicitly (but see Fig.~\ref{f:seq}).

We estimated the errors in the kinematic measurements by means of 
Monte-Carlo simulations. The parameters of the LOSVD were extracted
from a hundred realisations of the observed spectrum. Each pixel of a
Monte-Carlo spectrum was constructed adding a value randomly taken
from a Gaussian distribution with the mean of the observed spectrum
and standard deviation given by a robust-sigma estimate of the
residual of the fit to the observed spectrum. All realisations provide
a distribution of values from which $1\sigma$ confidence levels were
estimated. During the extraction of the kinematics for error
estimates, we switched off the penalisation of the pPXF method in
order to obtain the true (unbiased) scatter of the values (see
\citealt{2004PASP..116..138C} for a discussion).

\subsection{Distribution and kinematics of ionised gas}
\label{ss:ions}
NGC~2974 has previously been searched for the existence of
emission-line gas. \citet{1988ApJ...330..684K} reports the detection
of HI in a disc structure aligned with the optical isophotes. The
total mass of H{\small I} is estimated to be 8$\times10^{8} \Msun$,
rotating in a disc with an inclination of i $\approx$
55\degr. \citet{1993A&A...280..409B} detected H$\alpha$ emission
distributed in a flat structure along the major axis. Assuming a disc
geometry, the inferred inclination is $\approx$ 59\degr, and the total
mass of H{\small II} was estimated to be $\approx$
3$\times10^{4}\Msun$. Similar results are found by
\citet{1998A&AS..128...75P}. Deep optical ground-based imaging studies
suggested the existence of ``arm-like'' spiral structures, visible in
the filamentary distribution of ionised gas outside $\sim5\arcsec$
\citep{1992ApJ...387..484B,1993A&A...280..409B}.  The recent
high-resolution HST imaging in H$\alpha$+[N{\small II}] revealed the
presence of a gaseous two-arm spiral in the inner $\sim200$ pc, with a
total mass of $6.8\times 10^{4} \Msun$
(EGF03\nocite{2003MNRAS.345.1297E}).

The strongest emission line in the \SAURON spectra of NGC~2974 is the
[O{\small III}] doublet. There is also considerable emission in
H$\beta$ and some emission from the [N{\small I}] lines. Measurement
of the emission-line kinematics followed the extraction of the
absorption-line kinematics. For each spectrum in the data-cube we
performed three steps:

\begin{itemize}

\item[(i)] The pPXF method provided the model absorption spectrum that
           yielded the best fit to the spectral range with the
           emission lines ([O{\small III}], H$\beta$ and [N{\small
           I}]) excluded.

\item[(ii)] We then subtracted the model absorption spectrum from the
            original observed spectrum. This resulted in a ``pure
            emission-line'' spectrum which was used to extract the gas
            kinematics.

\item[(iii)] Each emission line was approximated with a Gaussian. The
             fit was performed simultaneously to the three lines of
             [O{\small III}] and H$\beta$, not using the mostly
             negligible [N{\small I}] doublet.

\end{itemize}

This procedure assumes that the velocity and velocity dispersion of
the different emission lines are equal. Performing the simultaneous
fit to the lines while allowing them to be kinematically independent
yields similar results (Sarzi et al. in preparation). Following
\citet{1989agna.book.....O} we assumed a 1:2.96 ratio for the
components of the [O{\small III}] doublet, while leaving the intensity
of the [O{\small III}] and H$\beta$ lines independent. The flux maps
of [O{\small III}] and H$\beta$ lines as well as the maps of the
[O{\small III}] emission-line mean velocity and velocity dispersion
are presented in Fig.~\ref{f:gas_maps}. The instrumental broadening of
108 $\kms$ was subtracted in quadrature from the emission-line
velocity dispersion map presented here and also used in
Section~\ref{s:em_gas}.

Both the H$\beta$ and [O{\small III}] emission lines are present over
the whole extent of the maps on Fig.~\ref{f:gas_maps}, but their
intensity drops off approximately exponentially with distance from the
centre. The [O{\small III}] emission is stronger over the entire
\SAURON field with the [O{\small III}] to H$\beta$ line-ratio being
$\approx$1.7. The shape of the distributions are very similar,
although H$\beta$ follows the stellar light isophotes more
precisely. The [O{\small III}] distribution shows departures from the
stellar isophotes in two roughly symmetric regions, positioned east
and west of the centre on Fig.~\ref{f:gas_maps}. The nature of these
dips in the [O{\small III}] flux are discussed in
Section~\ref{ss:bar}.

\subsection{Ground- and space-based imaging}
\label{ss:image}
In this study we used the existing ground- and space-based images of
NGC~2974. The already reduced wide-field ground-based \emph{I}-band
image of NGC~2974 was taken from \citet{1994A&AS..104..179G}, obtained
at the 1.0-m Jacobus Kapteyn Telescope (JKT). We also retrieved the
Wide Field and Planetary Camera 2 (WFPC2) association images of
NGC~2974 from the Hubble Space Telescope (HST) archive (Program ID
6822, PI Goudfrooij). The details of all imaging observations are
presented in Table~\ref{t:images}.

\begin{figure}
\begin{center}
   \includegraphics[width=\columnwidth]{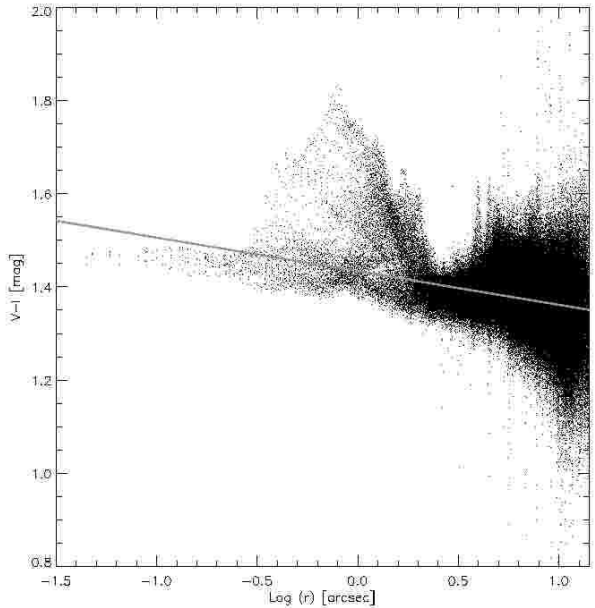}
   \caption{\label{f:colour} V-I colour versus the elliptical radius
   of every pixel in the inner 15\arcsec~of the WFPC2/PC1 images of
   NGC~2974. The straight line presents the best fit to the points
   obtained by minimising the absolute deviation. Notice the excess of
   red pixels between 0\farcs3 and 2\farcs5 caused by dust. }
\end{center}
\end{figure}

\begin{figure*}
\begin{center}
  \includegraphics[width=\textwidth]{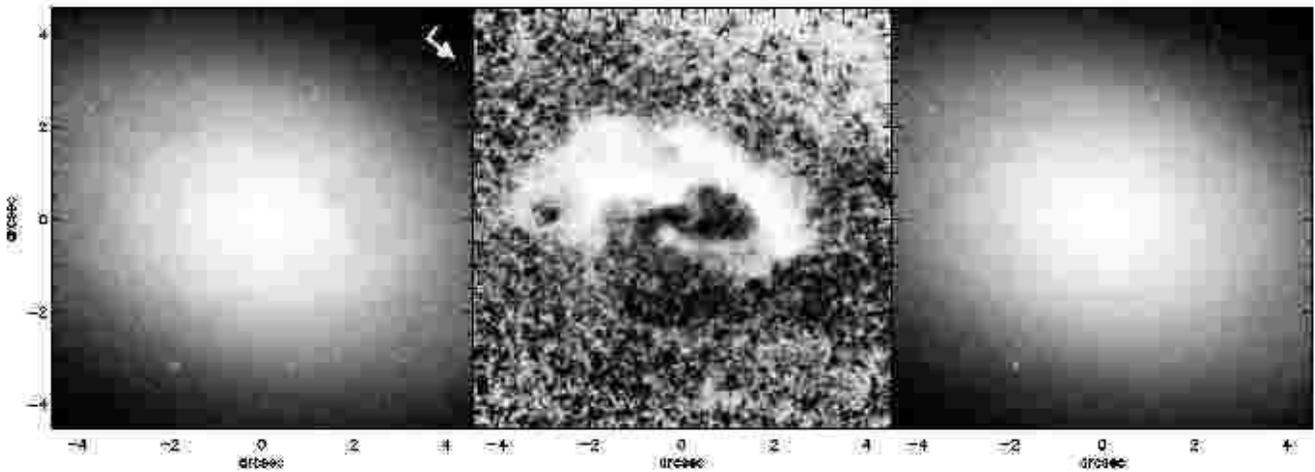}
  \caption{\label{f:excess} Dust correction on the F814W WFPC2 image
  of NGC~2974. From left to right: observed F814W image; colour excess
  E(V-I) obtained as described in the text; dust corrected F814W
  image. The arrow points to the north and associated dash to the
  east. All images were constructed using histogram
  equalisation. Lighter shades represent brighter regions. The E(V-I)
  map is stretched between -0.1 (dark) and 0.2 (bright) magnitudes.}
\end{center}
\end{figure*}

\begin{table}
 \caption[]{Summary of the ground- and space-based observations of NGC
 2974. The exposure times of the HST/WFPC2 observations are averages
 of all frames used to produce the WFPC2 association images.  }
  \label{t:images}
$$
  \begin{array}{lcc}
    \hline
    \noalign{\smallskip}
    & $JKT$ & $HST/WFPC2$\\
    \noalign{\smallskip}
    \hline
    \noalign{\smallskip}
    $Filter band$             & I            & $F$547$M$\, \&\, $F$814$W$\\
    $Exposure time (s)$       & 60           & 700\, \&\, 250 \\
    $Field of view (arcsec)$  & 380\times350 & 32\times32\\
    $Pixel scale (arcsec) $   & 0.3106       & 0.0455\\
    $Date of observations$    & 16.04.1993   & 16.04.1997\\
    \noalign{\smallskip}
    \hline
  \end{array}
$$
\end{table}

A major complication in the derivation of the surface brightness model
needed for the dynamical modelling is the existence of dust, clearly
visible on the high resolution images. We considered two possible
approaches: masking the patchy dust areas and excluding it from the
calculation of the model, or constructing a dust-corrected image. We
decided to adopt the latter approach to determine the stellar surface
brightness. We derived the correction of dust absorption using the
F547M and F814W WFPC2 images, following the steps listed in
\citet{2002ApJ...578..787C}. The process consists of construction of a
colour excess map E(V-I), from a calibrated V-I colour image. The
colour excess map is used to correct the pixels above a given E(V-I)
threshold using the standard Galactic extinction curve. We assumed
that the dust is a screen in front of the galaxy and that
dust-affected pixels have the same intrinsic colour as the surrounding
unaffected pixels. Figure~\ref{f:colour} shows the calibrated V-I
colour of pixels in the inner part of the PC images. The best fit to
the colours was obtained by minimising the absolute deviation of the
pixel values. This fit, represented by a line in Fig.~\ref{f:colour},
was used to calculate the colour excess by subtracting the measured
colour from the fit. The resulting E(V-I) image is shown in the second
panel of Fig.~\ref{f:excess}. The other panels on the same figure
present the inner parts of the F814W PC image before and after the
correction of dust absorption. The colour excess image highlights the
dust structure visible also on Fig.~3 of
EGF03\nocite{2003MNRAS.345.1297E} and suggests a non-uniform
distribution of dust in the central region of NGC~2974.

%
%

\section{Quantitative analysis of velocity maps}
\label{s:kin}
Two-dimensional kinematic maps offer a large amount of information and
are often superior to a few long-slit velocity profiles. The
two-dimensional nature of these data motivates us to quantify the
topology and structure of these kinematic maps, just as is commonly
done for simple imaging. We have developed a new technique to deal
with kinematic maps based on the Fourier expansion, and, due to its
similarity to the photometry, we named it kinemetry
\citep{2001sf2a.conf..289C}. This method is a generalisation of the
approach developed for two-dimensional radio data
\citep{1994ApJ...436..642F, 1997MNRAS.292..349S,
2004ApJ...605..183W}. The aim of the method is to extract general
properties from the kinematic maps of spheroidal systems (early-type
galaxies) without assuming a specific intrinsic geometry (e.g. thin
disc) for the distribution of stars. This changes the interpretation
and the approach to the terms of the harmonic expansion from the case
of cold neutral hydrogen or CO discussed in the above-mentioned
papers. In this section we briefly present the method and apply it to
the stellar and gaseous velocity maps (Krajnovi\'c et al. (in prep.)).

\subsection{Harmonic Expansion}
\label{ss:harm}
The kinemetry method consists of the straightforward Fourier expansion
of the line-of-sight kinematic property $K(r, \theta)$ in polar
coordinates:
\begin{equation}
  \label{eq:kinemetry}
  K(r,\theta) = a_{0}(r) +
                \sum_{n=1}^{N} c_{n}(r)\,\cos[n(\theta-\phi_{n}(r))].
\end{equation}
The expansion is done on a set of concentric circular rings (although
other choices are possible), and its main advantage is linearity at
constant $r$. The expansion is possible for all moments of the LOSVD,
but in this paper we restrict ourselves to the mean velocity maps.

The kinematic moments (moments of LOSVD) of triaxial galaxies in a
stationary configuration have different parity, e.g., mean velocity is
odd, while the second moment $\langle v{^2} \rangle$, is even. The
parity of a moment generates certain symmetries of the kinematics
maps. More generally, the maps of odd moments are
$point-anti-symmetric$, or:
\begin{equation}
  \label{eq:pas}
  V(r,\theta+\pi) = -V(r,\theta).
\end{equation}
If axisymmetry is assumed, in addition to the previous
relation, maps are $mirror-anti-symmetric$, or:
\begin{equation}
  \label{eq:mas}
  V(r,\pi - \theta) = -V(r,\theta),
\end{equation}
These symmetry conditions translate into the requirement on the
harmonic expansion (eq.~\ref{eq:kinemetry}) that for
point-anti-symmetric maps the even coefficients in the expansion are
equal to zero, while in the case of mirror-anti-symmetry,
additionally, the odd phase angles have a constant value, equal to the
photometric position angle (PA) of the galaxy in the case of a true
axisymmetric galaxy. This means that to reconstruct the mean velocity
map of a stationary triaxial galaxy, it is sufficient to use only odd
terms in the expansion.

These properties of the velocity maps enable certain natural filtering
(point-(anti)-symmetric --- eq.~(\ref{eq:pas}), and
mirror-(anti)-symmetric --- eq.~(\ref{eq:mas})) using the harmonic
expansion with coefficients set to zero or phase angles fixed at
certain values. For (visual) comparisons of the data with the results
of axisymmetric modelling it is useful to apply the axisymmetric
filtering to the data, as we will see below (Section~\ref{s:em_gas}).

\begin{figure}
\begin{center}
  \includegraphics[width=\columnwidth]{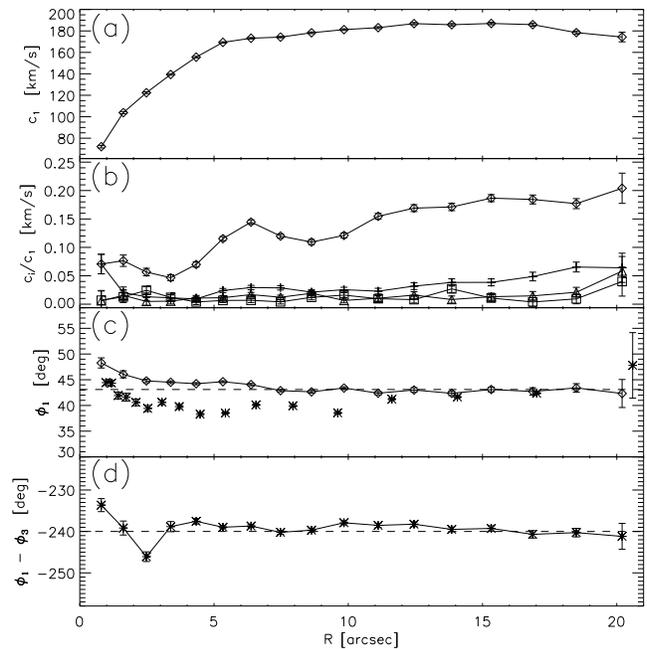}
  \caption{\label{f:stars_kin} Kinemetric expansion of the stellar
  velocity field as a function of radius. From top to bottom: (a)
  first amplitude coefficient in the harmonic expansion c$_{1}$; (b)
  ratios of amplitude coefficients presented by: triangles
  c$_{2}$/c$_{1}$, diamonds c$_{3}$/c$_{1}$, squares c$_{4}$/c$_{1}$
  and plusses c$_{5}$/c$_{1}$; (c) important position angles: phase
  angle $\phi_{1}$ representing the kinematic position angle
  (diamonds), photometric position angle as measured on WFPC2/PC1
  F814W image (asterisks) and adopted value for photometric position
  angle measured from the integrated \SAURON flux image (dashed
  straight line); (d) difference between the first and third phase
  from the kinemetric expansion.}
\end{center}
\end{figure}

\subsection{Kinemetric analysis of velocity maps}
\label{ss:a}
We wish to know the intrinsic shape of NGC~2974 and, in particular,
whether it is consistent with axisymmetry, which would permit the
construction of three-integral axisymmetric dynamical models of the
galaxy. In order to obtain the necessary information we applied the
kinemetric expansion to the observed stellar velocity map. If NGC~2974
is an axisymmetric galaxy, the kinemetric terms should have odd parity
(even terms should be zero) and the kinematic position angle should
be constant and equal to the PA.

\begin{figure}
\begin{center}
  \includegraphics[width=\columnwidth]{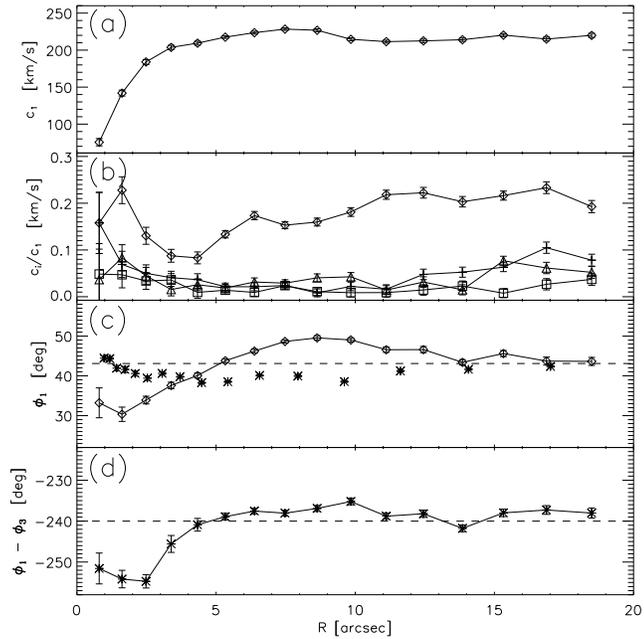}
  \caption{\label{f:gas_kin} Same as Fig~\ref{f:stars_kin}, but for
  the gas velocity map. }
\end{center}
\end{figure}

The amplitude and phases of the first five terms in the expansion are
presented in Fig.~\ref{f:stars_kin}. The first panel presents the
dominant term in the expansion, $c_{1}$, which gives the general shape
and amplitude of the stellar velocity map. The correction to this term
is given by the next significant term, $c_{3}$, which is already much
smaller than $c_{1}$ and is presented in the second panel as a
fraction of $c_{1}$. Even terms in the expansion, $c_{2}$ and $c_{4}$,
are also plotted on the same panel, and are much smaller ($\sim 1\%$
of $c_{1}$)\footnote{The zeroth term, $a_{0}$, gives the systemic
velocity of the galaxy and is not important for this analysis.}. For
comparison, the fifth term in the expansion, $c_{5}$, is also plotted
and is larger than both $c_{2}$ and $c_{4}$. The \SAURON~pixel size is
$\sim1\arcsec$ and measurement of higher-order terms at radii smaller
than 2\arcsec cannot be trusted. Clearly, the velocity map in NGC~2974
can be represented by the first two odd terms in the
expansion. Neglecting all higher terms results at most in a few
percent error.

The lower two panels in Fig.~\ref{f:stars_kin} present the phases of
the dominant terms. The $\phi_{1}$ phase is defined as the kinematic
angle of the velocity map, here measured east of north. This angle is
compared with measurements of two important angles: (i) the PA
measured on the WFPC2/PC F814W dust-corrected image using the {\tt
IRAF} ellipse fitting task {\tt ellipse} and (ii) the PA measured on
the reconstructed \SAURON flux image obtained by integrating the
spectra in each bin. The agreement between the different angles
measured on the \SAURON observations is excellent, with slight
departures in the inner 3\arcsec. The PA measured on the high
resolution WFPC2 image suggests a small photometric twist in the inner
10\arcsec~of $\approx 3\degr$, which can be also seen in
Fig.~\ref{f:mge}.

The phase angle $\phi_{3}$ is the phase of the third term in the
kinemetric extraction. It is easy to show, if the galaxy is
axisymmetric (requiring in eq.~\ref{eq:kinemetry}, K(r,$\theta$) = 0
for $\theta = \phi_{1} + \pi/2$), and the higher terms can be
neglected, that the phases $\phi_{1}$ and $\phi_{3}$ satisfy the
relation:
\begin{equation}
  \label{eq:axi_phase}
  \phi_{1} - \phi_{3} = \frac{n\pi}{3}
\end{equation}
where $n \in \mathbb{Z}$. The last panel in Fig.~\ref{f:stars_kin}
shows this phase difference. The condition given by
eq.(~\ref{eq:axi_phase}) is satisfied along the entire investigated
range with a small deviation in the inner 3\arcsec. Summarising all
the above evidence, we conclude that the observed stellar kinematics
in NGC~2974 is consistent with axisymmetry.

We repeated the kinemetric analysis\footnote{A similar analysis
approach for a gas disc would be using the \citet{1997MNRAS.292..349S}
harmonic analysis on a tilted-ring model of the gas disc, interpreting
the results within epicycle theory \citep[see
also][]{2004ApJ...605..183W}.} on the emission-line gas velocity maps
and present the results in Fig.~\ref{f:gas_kin}. While the amplitude
coefficients, $c_{i}$, are similar to the stellar coefficients (small
values of all terms higher than $c_{3}$), the behaviour of the phase
angles is quite different.

The last panel of Fig.~\ref{f:gas_kin} is perhaps the best diagnostic
tool. The dashed line presents the required value for the difference
between the phase angles, $\phi_{1} - \phi_{3}$, assuming
axisymmetry. Deviations are present in the inner $4\arcsec$ and,
although much smaller, between $9\arcsec$ and $11\arcsec$.  These
deviations indicate departures from axisymmetry, which are strongest
in the central few arcsecs.

\begin{figure}
        \includegraphics[width=\columnwidth]{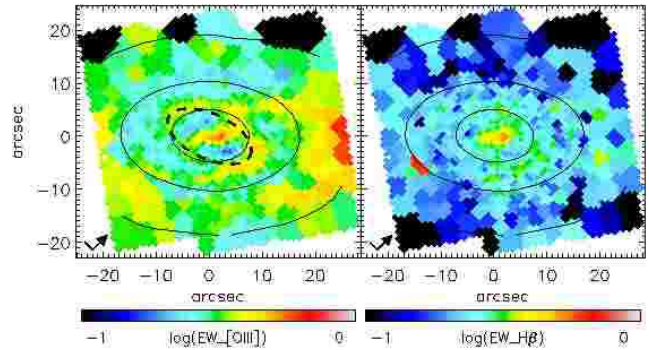}
        \caption{\label{f:ew} Equivalent widths of [OIII] and H$\beta$
        emission-lines. These plots, constructed by dividing the
        emission-line flux by the symmetrised stellar continuum, show
        the relative distribution of emission-lines to the stellar
        continuum.  The dashed ellipse represents the possible OLR
        ring of the inner bar (see text for details). Overplotted
        contours are levels of the reconstructed total intensity from
        the full \SAURON spectra. The arrow in the left corner of each
        plot marks North-East orientation of the maps. }
\end{figure}

\subsection{Signature of bars in NGC~2974}
\label{ss:bar}
In the previous section, we quantified the signatures of
non-axisymmetry on the gas velocity maps, especially strong in the
inner $4\arcsec$. Inside this radius,
EGF03\nocite{2003MNRAS.345.1297E} discovered a two-armed spiral and
explained it by a weak bar with the corotation resonance (CR) at
4\farcs9 and outer-Lindblad resonance (OLR) at 8\farcs5. These scales
are consistent with the observed departures from axisymmetry in the
\SAURON kinematic maps. An additional confirmation comes from the
equivalent width maps of emission-lines (Fig.~\ref{f:ew}). These maps
show the distribution of the emission-lines relative to the stellar
continuum. Especially noticeable is the ring structure in the
[O{\small III}] equivalent width distribution. For comparison, we
overplotted an ellipse with semi-major axis length of 8\farcs5 (the
radius of OLR of the EGF03\nocite{2003MNRAS.345.1297E} bar). The
orientation ($\sim 25\degr$ away from the major axis of the galaxy)
and the size of the ring ($\sim 8\farcs5$) are in good agreement with
the expected characteristics of the bar discovered by
EGF03\nocite{2003MNRAS.345.1297E}.

The equivalent width maps show that [O{\small III}] emission-line is
also influenced outside the OLR. There is an increase in the value of
the equivalent width at larger radii as well as a filamentary (spiral)
structure connecting the ring and this large scale region. Also, as
seen before on Fig.~\ref{f:gas_maps}, the distribution of the
[O{\small III}] emission-line intensity exhibits an elongated
structure in the central 5\arcsec. Similarly, around $8\arcsec -
10\arcsec$ [O{\small III}] is also elongated, but this time
approximately perpendicular to the first elongated
structure. Following this, the [O{\small III}] map has a plateau
between 12\arcsec\ and 15\arcsec. The end of the plateau is followed
by a dip in the distribution with a possible turn up at radii larger
than 20\arcsec.

Although the central structure and kinematics of the [O{\small III}]
distribution are influenced by the inner bar, the large-scale
structure (beyond $\approx10\arcsec$) is not likely to be influenced
by this weak inner bar. On a more speculative basis, we can infer the
existence of a large-scale bar. Studies of double bars
\citep{2002AJ....124...65E,2002ApJ...567...97L,2003ApJS..146..299E}
suggest a size ratio of about 5 to 10 between the primary and
secondary bars; this would set the size of an hypothetical large-scale
(primary) bar in NGC~2974 to between about $12\farcs5$ and
$25\arcsec$.

\begin{figure}
        \includegraphics[width=\columnwidth]{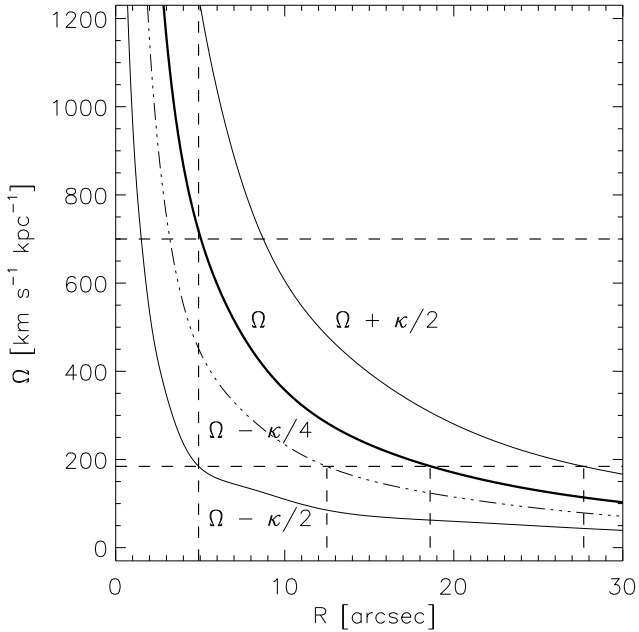}
        \caption{\label{f:res_dia} Diagram of resonances in NGC~2974,
        derived from the potential of the best-fitting stellar
        dynamical model. The upper horizontal line is the pattern
        speed of the nuclear (secondary) bar from the EGF03 study
        ($\Omega^s_p=700 \kms$ kpc$^{-1}$). The lower horizontal line
        shows the position of the inferred pattern speed of the
        large-scale primary bar ($\Omega^p_p=185 \kms kpc^{-1}$). The
        vertical lines show the assumed radial positions of important
        resonances of the primary bar. From left to right: ILR (CR of
        the secondary bar), UHR, CR and OLR. }
\end{figure}

A more precise, although still approximate, estimate of the properties
of the primary bar can be obtained from the resonance curves of
NGC~2974. Using the potential of the best-fitting stellar dynamical
model (see Section~\ref{s:dyno}) we constructed the resonance diagram
presented in Fig.~\ref{f:res_dia}. We calculated profiles of $\Omega$,
$\Omega - \kappa/2$, $\Omega + \kappa/2$ and $\Omega - \kappa/4$,
where $\Omega$ is the angular velocity, $V/r$, and $\kappa$ is
epicyclic frequency, defined as $\kappa^2 = r\frac{\ud \Omega^2}{\ud
r} + 4\Omega^2$.  Assuming that the resonances of the primary and the
secondary bar are coupled to minimise the chaos produced around the
resonances \citep{1993A&A...277...27F, 1990ApJ...363..391P}, where the
inner Lindblad resonance (ILR) of the primary bar is the CR of the
secondary bar, we can estimate the main resonances of the primary bar
(indicated by the vertical lines on Fig.~\ref{f:res_dia}). The pattern
speed of the primary component is then $\approx 185\kms$ kpc$^{-1}$,
ILR is at $4\farcs9$, ultra harmonic resonance (UHR) at $\approx
12\farcs5$, CR at $\approx 18\farcs6$ and OLR at $\approx
27\farcs7$. The size of the primary bar can be taken to be 80\% of its
CR, so about 13 to 14\arcsec ($\sim 1.4$~kpc).

The presented analysis is strictly valid only for an axisymmetric
potential with a weak bar perturbation, hence the above-mentioned size
estimates are only approximate, but indicative. We see several
features in the [O{\small III}] distribution and velocity maps that
support the assumption of a large-scale primary bar: the dip in the
gas velocity map around 12\arcsec, the plateau between 12\arcsec and
15\arcsec in the [O{\small III}] distribution as well as the dip in
the [O{\small III}] distribution around 18\arcsec. The last one
corresponds to the position of the CR, which is a chaotic region
devoid of gas, consistent with the observed lower flux in that region.

\subsection{Case for axisymmetry in NGC~2974}
\label{ss:axisym}

\begin{figure}
\begin{center}
  \includegraphics[width=\columnwidth]{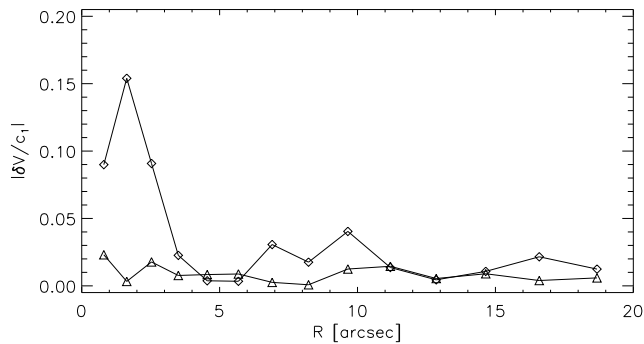}
  \caption{\label{f:deltaV_c1} Contribution of the non-axisymmetric
  motion to the stellar and gaseous velocity fields, as a fraction of
  the dominant term in the kinemetric expansion, $c_{1}$. Diamonds
  represent the emission-line gas velocity contribution, while
  triangles denote the stellar velocity contribution. }
\end{center}
\end{figure}

The alignment of the gaseous and stellar component, previously
detected and also confirmed in this study, suggests that NGC~2974 is
an axisymmetric galaxy. However, the gaseous component shows
signatures of non-axisymmetric perturbations. The contribution of the
non-axisymmetric motion, $\Delta V/c_{1}$, to the total velocity field
can be quantified from the phase difference $\phi_{3}-\phi_{1}$. If
the condition in eq.~(\ref{eq:axi_phase}) is not satisfied then:
\begin{equation}
  \label{eq:non_axi}
  \frac{\Delta V}{c_{1}} = \frac{c_{3}}{c_{1}} \sin3(\phi_{1} -
  \phi_{3}),
\end{equation}
which is presented in Fig.~\ref{f:deltaV_c1} for both stellar and
gaseous velocity maps. At $\sim3\arcsec$, $\Delta V/c_{1}$ for the
emission-line gas is $\approx0.1$ and at $\sim10\arcsec$ it is
$\approx0.04$, confirming that the non-axisymmetric contribution is
significant in the centre of the emission-line velocity map. Its
influence on the stellar velocity is not significant over the \SAURON\
field. Emission-line gas is a more responsive medium and unlike the
stars, due to the viscosity of the gas particles, shows evidence of
weak non-axisymmetric perturbations. It is possible that other
early-type galaxies with disc-like components harbour such weak and
hidden bar systems.

Summarising, the stellar velocity map is point- and
mirror-anti-symmetric, supporting an axisymmetric shape for
NGC~2974. On the other hand, the gaseous velocity map shows strong
deviations from mirror-anti-symmetry in the centre, and the
distribution and equivalent width of [O{\small III}] emission lines
supports the weak inner bar found by EGF03\nocite{2003MNRAS.345.1297E}
and suggests the existence of a weak large-scale bar. However, since
the bar perturbations on the axisymmetric potential are presumed to be
weak and appear do not influence the stellar kinematics, we ignore
them in the remainder of the paper, and describe NGC~2974 with an
axisymmetric potential.

%
%

\section{Stellar Dynamical Modelling}
\label{s:dyno}
In order to investigate the orbital structure of NGC~2974 we construct
fully general axisymmetric models of the galaxy's stellar
component. The three-integral models presented here are based on
Schwarzschild's orbit superposition method
\citep{1979ApJ...232..236S,1982ApJ...263..599S}, further developed by
\citet{1997ApJ...488..702R}, \citet{1998ApJ...493..613V} and
\citet{1999ApJS..124..383C}, and adapted for more general
surface-brightness distributions by \citet[][hereafter
C02]{2002ApJ...578..787C} and \citet[][hereafter
V02]{2002MNRAS.335..517V}, similarly as in
\citet{1999ApJ...514..704C}. The three-integral modelling technique is
widely used for constructing dynamical models of axisymmetric
galaxies, and has been thoroughly described in the literature by the
aforementioned authors as well as by other groups
\citep[e.g.][]{2003ApJ...583...92G,2004ApJ...602...66V}. It is most
commonly used to determine the masses of the central black holes in
nearby galaxies and investigate the internal orbital structure of the
galaxies. The \SAURON observations of NGC~2974 do not have the
necessary resolution to probe the sphere of influence of the central
black hole and we therefore restrict ourselves to the determination of
the mass-to-light ratio $\Upsilon$ and the inclination $i$ of the
galaxy, as well as the internal orbital structure.

\subsection{The Multi-Gaussian expansion mass model}
\label{ss:mass}
The starting point of the stellar dynamical modelling is the
determination of the gravitational potential of the galaxy. The
potential can be obtained by solving the Poisson equation for a given
density distribution which can be derived by deprojecting the
observations of the 2D stellar surface density. In this work we used
the multi-Gaussian expansion (MGE) method \citep{1994A&A...285..723E},
following the approach of C02 and V02.

\begin{table}
   \caption{The MGE parameters of the circular PSF of HST/WFPC2/F814W filter.}
   \label{t:PSF_fit}
$$
   \begin{array}{ccc}
       \hline
       \noalign{\smallskip}
       $j$ & $G$_{j}  & \sigma_{k} $(arcsec)$ \\
       \noalign{\smallskip}
       \hline
            1 & 0.352 & 0.024\\
            2 & 0.531 & 0.072\\
            3 & 0.082 & 0.365\\
            4 & 0.035 & 0.908\\
       \noalign{\smallskip}
       \hline
   \end{array}
$$
\end{table}

\begin{table}
    \caption{The parameters of the MGE model of the deconvolved I-band
   WFPC2 surface brightness of NGC~2974. Columns present (from left to
   right): number of the two-dimensional Gaussian, central intensity
   of the Gaussian, width (sigma) of the Gaussian, axial ratio of the
   Gaussian, total intensity contained in the Gaussian.}
    \label{t:NGC2974_MGEfit}
$$
   \begin{array}{ccccc}
       \hline
       \noalign{\smallskip}
       $j$ & $I$_{j} (L_{\odot}{pc}^{-2}) &\sigma_{j} $(arcsec)$ &$q$_{j} & $L$_{j} (\times10^{9} L_{\odot}) \\
       \noalign{\smallskip}
       \hline
 1 & 190297.  &    0.0378596  &     0.580000   &   0.0108\\
 2 & 44170.6  &    0.0945030  &     0.800000   &   0.0215\\
 3 & 24330.8  &     0.185143  &     0.800000   &   0.0455\\
 4 & 27496.3  &     0.340087  &     0.583279   &   0.1264\\
 5 & 23040.6  &     0.591227  &     0.720063   &   0.3952\\
 6 & 10299.6  &      1.15500  &     0.777448   &   0.7279\\
 7 & 5116.29  &      3.41758  &     0.658664   &   2.6820\\
 8 & 1902.25  &      8.67562  &     0.597636   &   5.8305\\
 9 & 388.278  &      17.5245  &     0.677645   &   5.5060\\
10 & 139.447  &      43.9864  &     0.580000   &  10.663\\
11 & 16.9405  &      82.9488  &     0.800000   &   6.3538\\
       \noalign{\smallskip}
       \hline
   \end{array}
$$
\end{table}

In order to get the MGE model, we simultaneously fitted the
ground-based I-band image and the dust-corrected PC part of the
WFPC2/F814W image using the method and software developed by
\citet{2002MNRAS.333..400C}.  The dust correction
(Section~\ref{ss:image}) successfully removed the dust contamination
from the high-resolution image of the nucleus, but the large-scale
image was badly polluted by several stars, with a particularly bright
one almost on the galaxy's major axis. We masked all stars inside the
model area to exclude them from the fit. The ground-based image, used
to constrain the fit outside 25\arcsec, was scaled to the WFPC2/PC1
image. We computed the PSF of the F814W PC1 image at the position of
the nucleus of NGC~2974, using the TinyTim software (Krist \& Hook 2001),
and parametrised it by fitting a circular MGE model with constant
position angle as in C02.  Table~\ref{t:PSF_fit} presents the relative
weights G$_{j}$ (normalised such that their sum is equal to unity) and
the corresponding dispersions $\sigma$ of the four Gaussians.
Table~\ref{t:NGC2974_MGEfit} gives the parameters of the MGE model
analytically deconvolved from the PSF. Following the prescription of
\citet{2002MNRAS.333..400C}, we increased the minimum axial ratio of
the Gaussians, \emph{q$_{j}$}, until the $\chi^{2}$ significantly
changed, in order to make as large as possible the range of allowed
inclinations by the MGE model. The upper limit to the \emph{q$_{j}$}
was also constrained such that the MGE model is as close as possible
to a density stratified on similar ellipsoids. Although the
deprojection of an axisymmetric density distribution is non-unique
\citep{1987IAUS..127..397R}, our `regularisation' on the MGE model
produce realistic intrinsic densities, while preventing sharp
variations, unless they are required to fit the surface brightness. We
verified that the MGE model used in this study is consistent with the
MGE model presented in EGF03\nocite{2003MNRAS.345.1297E}.

\begin{figure}
\begin{center}
   \includegraphics[width=8.cm]{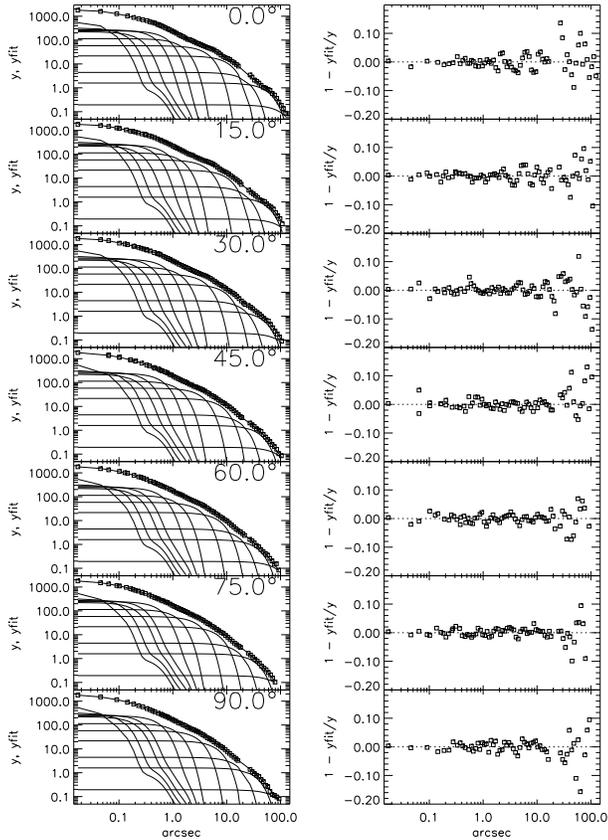}
   \caption{\label{f:prof} {\it Left Panels:} Comparison between the
    combined WFPC2/F814W and ground-based photometry of NGC~2974 (open
    squares) and the MGE model (solid line) in seven angular sectors
    as function of radius. The individual convolved Gaussians are also
    shown. \textit{Right panels:} radial variation of the relative
    error along the profiles.}
\end{center}
\end{figure}

\begin{figure}
\begin{center}
  \includegraphics[width=8.cm]{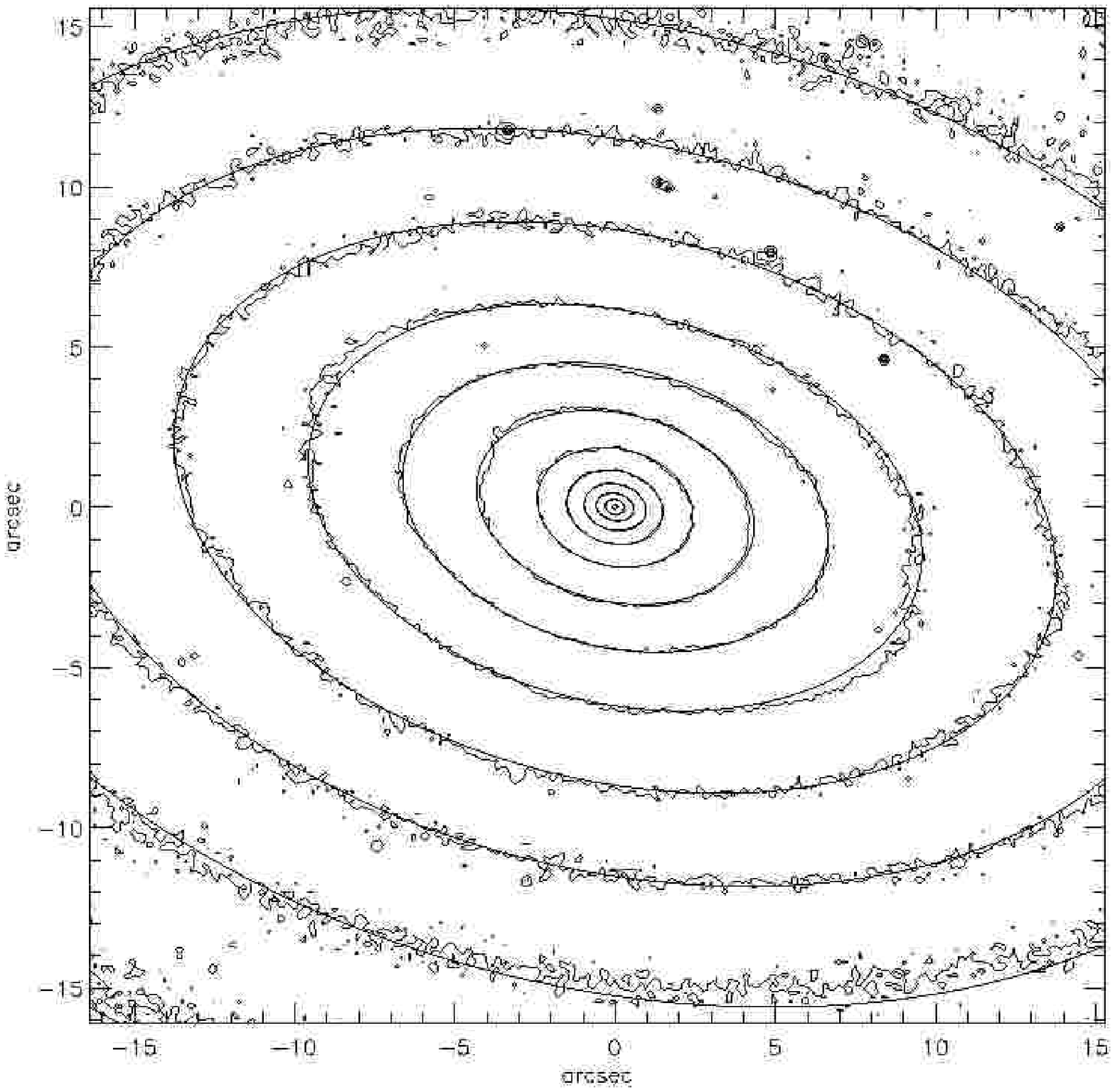}
  \includegraphics[width=8.cm]{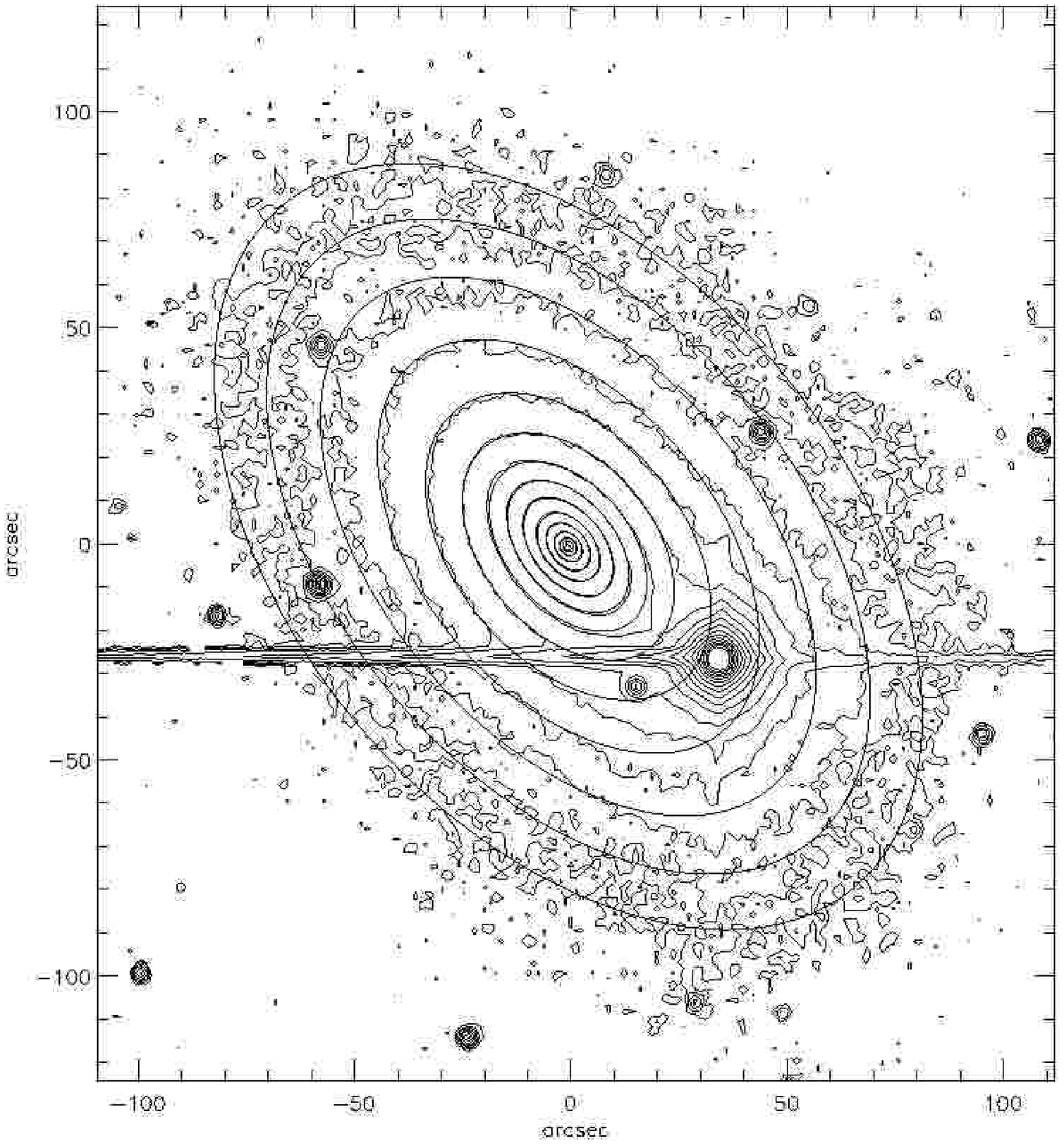}
  \caption{ \label{f:mge} Contour maps of the ground-based I-band and
  dust-corrected WFPC2/F814W images. The brightest star on the
  ground-based image as well as four additional stars covered by the
  model were masked out and excluded from the fit. Superposed on the
  two plots are the contours of the MGE surface brightness model,
  convolved with the WFPC2 PSF.}
\end{center}
\end{figure}

\begin{figure}
\begin{center}
  \includegraphics[width=\columnwidth]{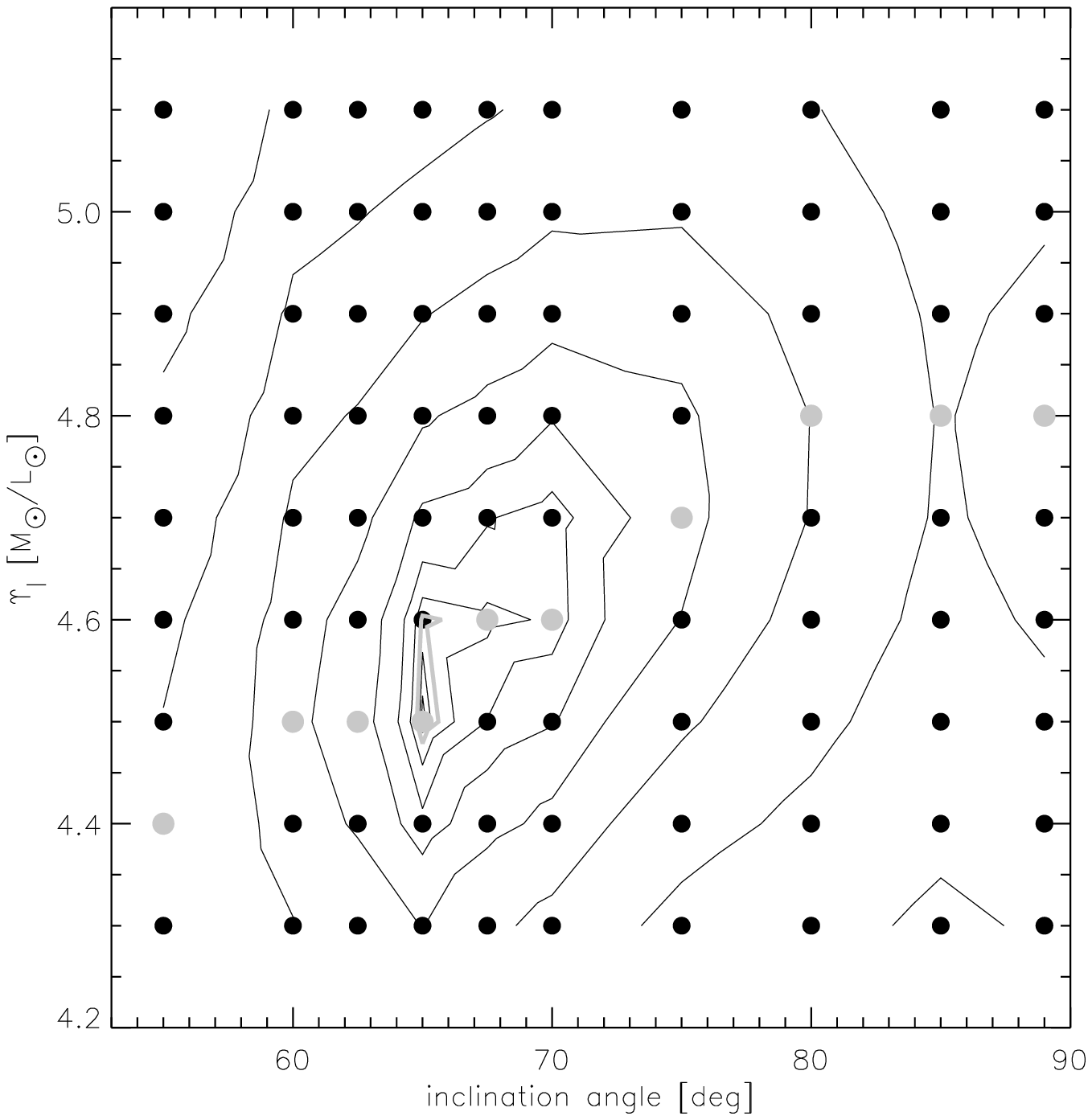}
  \caption{\label{f:grid} A grid of inclination angle $i$ versus
  $M/L$ ratio $\Upsilon$. Contours present constant $\Delta \chi^{2}$,
  measuring the goodness of fit of the dynamical models. Every symbol
  corresponds to a three-integral axisymmetric model with given
  inclination angle and $\Upsilon$ ratio. The best-fitting model for
  each inclinations are presented with grey symbols. The first three
  contours show the formal 68.3\%, 95.4\% and 99.73\% (thick grey
  contour) confidence regions for two degrees of freedom. The
  best-fitting model on the grid has an inclination of 65\degr and
  $\Upsilon$ = 4.5. }
\end{center}
\end{figure}

The comparison between the MGE model and the photometry along
different angular sectors is shown in Fig.~\ref{f:prof}. The profiles
are reproduced to within 3\%, and the RMS error is about 2\%. The
increase of the relative error (right-hand panel in
Fig.~\ref{f:prof}) at larger radii is caused by the light pollution of
the bright star on the major axis of the galaxy and can be better
understood by looking at the comparison of the convolved model and the
actual observation. Fig.~\ref{f:mge} presents both ground- and
space-based images and the MGE model. On the WFPC2/PC image there is a
slight deviation from the model (of constant PA) about 10\arcsec~from
the nucleus. The deviating structure is point-symmetric and
reminiscent of a spiral perturbation. as suggested by other studies
(\citealt{1992ApJ...387..484B}; \citealt{1993A&A...280..409B};
EGF03\nocite{2003MNRAS.345.1297E}). Except for this slight departure
from the model the galaxy surface brightness is well represented by
the MGE model and we use it to calculate the representative
gravitational potential.

\subsection{Construction of three-integral models}
\label{ss:3I}
Briefly, Schwarzschild's method can be divided in four steps. In the
first step, one derives the stellar potential assuming the shape of
the potential (axisymmetry) and stellar mass-to-light ratio,
$\Upsilon$ (free parameter), through deprojection of a parametrisation
of the surface density (in this case MGE parametrisation which can be
deprojected analytically). The second step involves the construction
of a representative orbit library by integrating the orbits in the
derived potential. Each orbit is specified by three integrals of
motion ($E, L_{z}, I_{3}$), where $E$ is the energy, $L_{z}$ the
component of the angular momentum along the $z$ symmetry axis, and
$I_{3}$ is a non-classical integral, which is not known
analytically. The integral space is constructed on a grid that
includes $>$99\% of the total luminous mass of the galaxy. The next
step consists of projecting the orbits onto the space of observables
($x'$, $y'$, $v_{los}$), where $(x', y')$ are in the plane of the sky
and $v_{los}$ is the line-of-sight velocity given by the
observations. In our implementation, this is done taking into account
the PSF convolution and aperture binning. The final step of the method
is to determine the set of weights for each orbit that, when added
together, best corresponds to the observed kinematics in the given
spatial bin as well as reproducing the stellar density. In our
implementation of the method, this best-fitting set is found by
solving a non-negative least-squares (NNLS) problem using the routine
written by \citet{1974slsp.book.....L}.

The software implementation used here is similar to that used in the
V02\nocite{2002ApJ...578..787C} and C02\nocite{2002MNRAS.335..517V}
studies, but it has evolved substantially since. The improvements are
described in detail in Cappellari et al.\ (in prep.). We verified
that the results from the new code are the same as from the old one if
identical settings are adopted. An application to the elliptical
galaxy NGC~4473 was presented in \citet{2004cbhg.sympE...5C}. Here we
give a quick overview of the changes with respect to the description
in C02:

\begin{enumerate}

\item[1.] The method requires that the orbits sample a
          three-dimensional space of integrals of motion, the energy
          $E$, $L_{z}$ and $I_{3}$. In the new scheme (see
          \citealt{1999ApJS..124..383C} for details of the previous
          approach), at each $E$, we construct a polar grid of initial
          starting positions on the meridional plane (linear in angle
          and in radius), going from $R=z=0$ to the curve defined by
          the thin tube orbits (to avoid duplication) which is well
          approximated by the equation $R^{2}+z^{2}=R^{2}_{c}(E)$,
          where $R_{c}(E)$ is the radius of the circular orbit at
          energy E. The orbits are released with $v_{R}=v_{z} = 0$ and
          $L_{z}\ne 0$. In this way we sample the observable space by
          uniformly distributing the position of the orbital cusps
          (see \citealt{2004cbhg.sympE...5C}) on the sky plane.

\item[2.] Improved treatment of seeing effects and instrumental
          point spread function (PSF) by a Monte Carlo method. The PSF
          can be considered as the probability that an observed photon
          arriving at a detector will be displaced from its original
          position by a given amount (specified by the PSF
          characteristics). The projected orbital points (results of
          the orbit integration and projection onto the sky-plane)
          are stored in the apertures in which they landed after
          applying a random displacement taken from the Gaussian
          probability distribution defined by the PSF.

\item[3.] Generalisation of the projection of the orbits into the
          space of the observables. The bins of the optimal
          Voronoi binning of the two-dimensional integral-field data
          have non-rectangular shapes. The orbital observables now can
          be stored on apertures of any shape that can be represented
          by polygons.

\end{enumerate}

\subsection{Stellar dynamics - modelling results and discussion}
\label{ss:model}
Our stellar kinematic maps of NGC~2974 consist of 708 Voronoi
bins. Each bin contributes with 6 kinematic observables to which we
also add the intrinsic and projected mass density observables,
resulting in a grand total of 5664 observables. The largest orbit
library that was computationally possible for the given number of
observables, consists of $2\times41\times10\times10 = 8200$ orbits
(for each of the 41 different $E$ we construct a polar grid of
starting points sampling 10 angles and 10 radii). With this choice of
orbit library, the number of observables is smaller than the number of
orbits, and the NNLS fit will not have a formally unique
solution. Moreover the recovery of the orbital weights for the orbits
from the observations is an inverse problem, and as such is
intrinsically ill-conditioned. For these reasons a direct solution of
the problem generally consists exclusively of sharp isolated peaks. It
is unlikely for the DF of real galaxies to be very jagged, since
(violent) relaxation processes tend to smooth the DF. Moreover
observational constraints on the smoothness of the DF, at least for
the bulk of the stars in a galaxy, come from the smoothness of the
observed surface brightness, down to the smallest spatial scales
sampled by HST.

\begin{figure*}
\begin{center}
  \includegraphics[width=\textwidth]{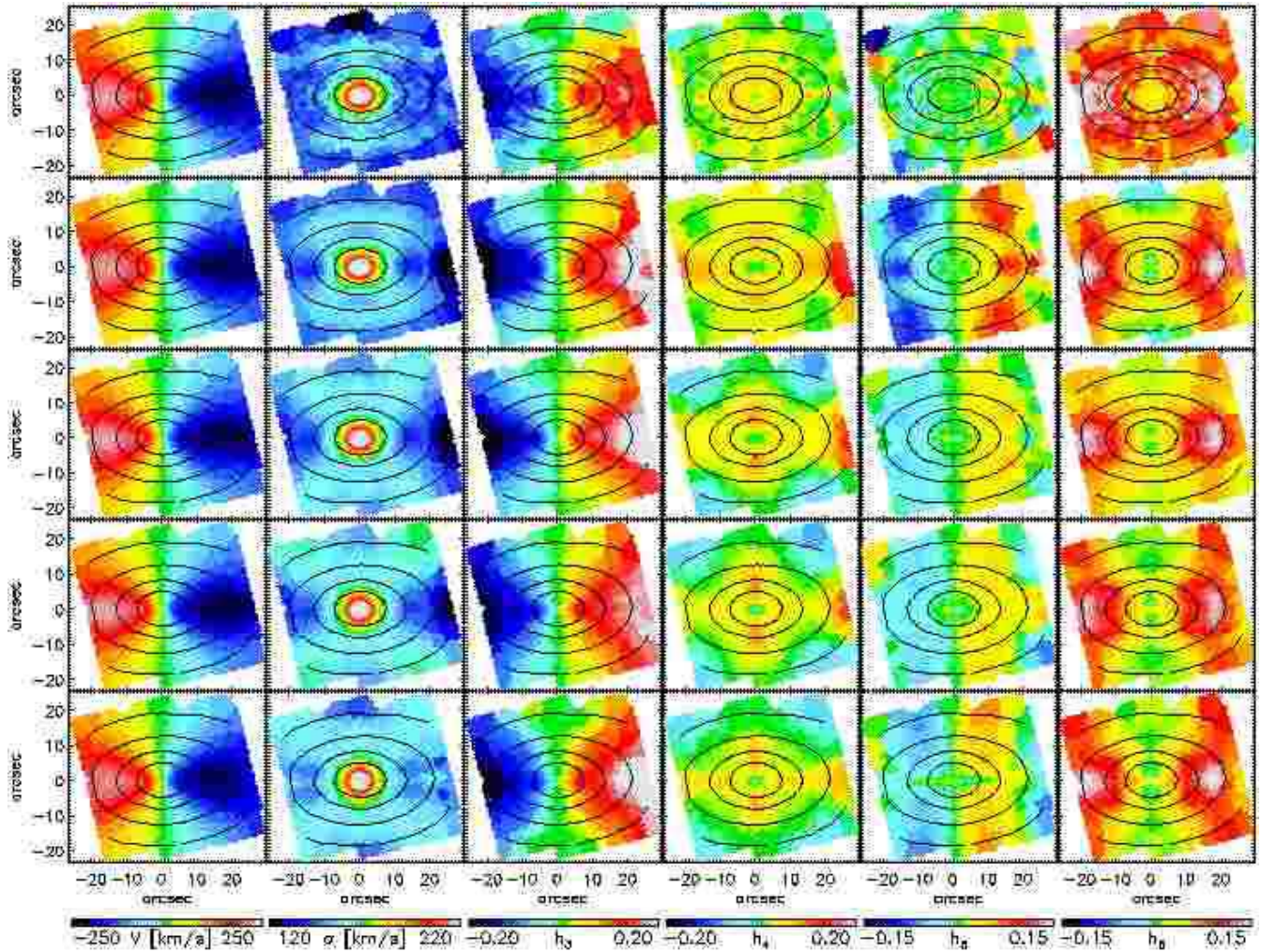}
  \caption{\label{f:seq} Comparison of the symmetrised observation of
  NGC~2974 (first row) and four orbital superposition models with best
  mass-to-light ratio $\Upsilon$, for a given inclination
  \emph{i}. From second row to bottom: ($i,\Upsilon$) = ($55,4.4$),
  ($65,4.5$), ($75,4.7$), ($89,4.8$). From left to right each panel
  presents: mean velocity $V$, velocity dispersion $\sigma$, and
  Gauss-Hermite moments: $h_{3}, h_{4}, h_{5}$ and $h_{6}$. Isophotal
  contours of total light are shown with solid lines.}
\end{center}
\end{figure*}

\begin{figure*}
\begin{center}
  \includegraphics[width=\textwidth]{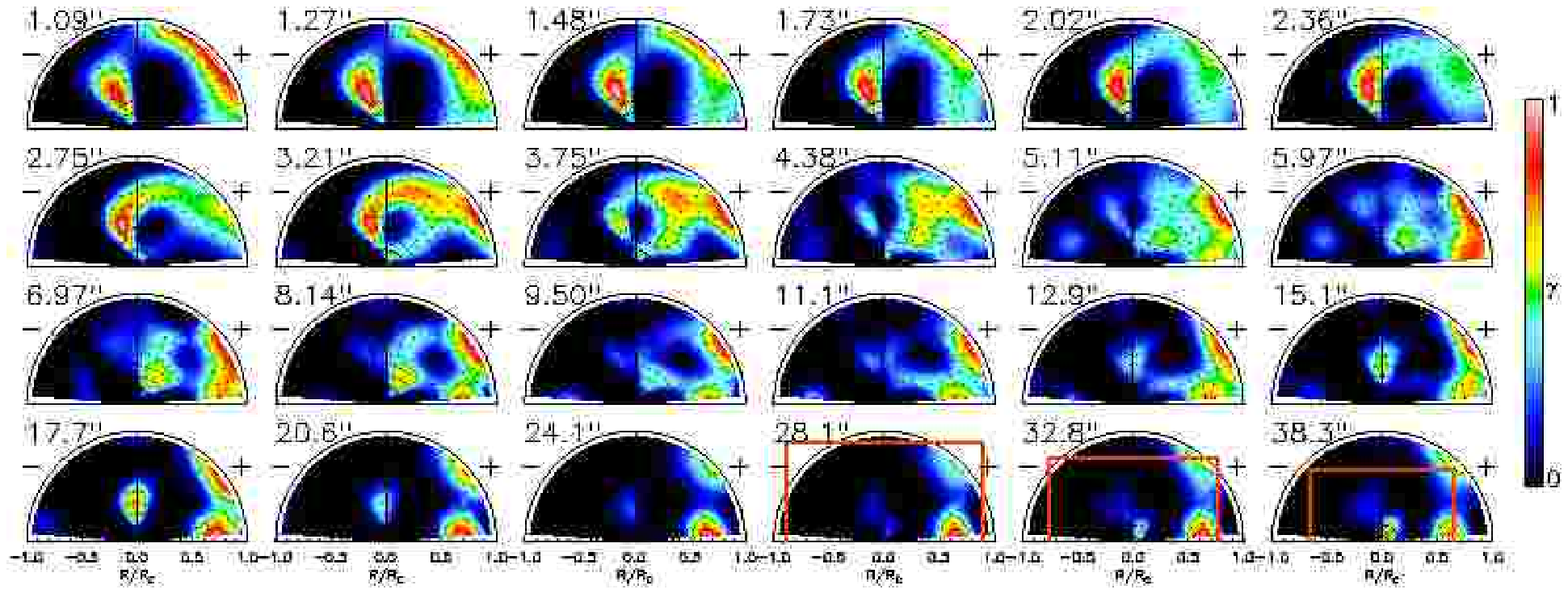}
  \caption{\label{f:int_space} Integral space of the best-fitting
  orbit superposition model for NGC~2974. Each panel plots the
  meridional plane ($R$,$z$) with the starting positions of orbits
  (dots) for the given energy. Orbital starting positions correspond
  to the position of cusps ($v_{R} = v_{z} = 0$). Overplotted is the
  fraction of mass assigned to orbits at constant energy, labelled by
  the radius of the circular orbit in arcseconds (printed at the upper
  left corner of each panel). The radius of the circular orbit is the
  size of the horizontal axis (measured from the centre). Negative
  values are obtained by sign reversal, since orbits can be prograde
  and retrograde. Orbits with high angular momentum are found in the
  right and left corner respectively on the plots. We show only the
  radii constrained by the data. The last three panels have the size
  of the \SAURON\ field overplotted for comparison (red
  rectangle). The area inside the rectangle is constrained by the
  kinematic data. The colour bar on the right represents the relative
  colour coding of the orbital mass weights ($\gamma$), where 1
  represent the largest mass weight assigned to an orbit at the given
  energy. }
\end{center}
\end{figure*}
%

A standard mathematical approach to solve inverse problems is by
regularising \citep[e.g.][Chapter 18]{pre92}. This has been generally
applied by all groups involved in this modelling approach
\citep[e.g.][]{1997ApJ...488..702R, 2003ApJ...583...92G,
2004ApJ...602...66V, 2004MNRAS.347L..31C}. Regularisation inevitably
biases the solution, by forcing most orbits to have a non-zero
weight. The key here is to apply the right amount of
regularisation. Previous tests with the Schwarzschild code suggested a
value of the regularisation parameter $\Delta=4$ (see
\citealt{1999ApJS..124..383C}; V02\nocite{2002MNRAS.335..517V}). After
initial testing we also adopted this value. For more details on
regularisation, see McDermid et al.\ (in prep.).

The models are axisymmetric by construction. In order to avoid
possible systematic effects we additionally symmetrise the stellar
kinematics as usually done in other studies. The symmetrisation uses
the mirror-(anti)-symmetry of the kinematic fields, such that
kinematic values from four symmetric positions ($(x,y), (x,-y),
(-x,y), (-x,-y)$) were averaged. However, the Voronoi bins have
irregular shapes and they are not equally distributed with respect to
the symmetry axes of the galaxy (minor and major axes). In practice,
we average the four symmetric points, and, if for a given bin there
are no bins on the symmetric positions, we interpolate the values on
those positions and then average them. As the number of data points is
not decreased in this way, the errors were left unchanged.

Finally, we fitted axisymmetric dynamical models to symmetrised
observations of velocity, velocity dispersion and Gauss-Hermite
moments ($h_{3}-h_{6}$) while varying the mass-to-light ratio
$\Upsilon$ and the inclination $i$. Figure~\ref{f:grid} shows a grid
of our models with overplotted $\Delta \chi^2$ contours. The best-fit
parameters are $\Upsilon = 4.5 \pm 0.1$ (in the $I$ band) and
$i=65\degr \pm 2.5\degr$.  The data -- model comparison for these
values is given in Fig.~\ref{f:seq} (symmetrised data are shown in the
first row and the best-fitting model in the third row). In the same
figure we present the best-fitting models for the given
inclinations. The formal statistical analysis firmly rules out (with
$3\sigma$ confidence) all inclinations outside $i = 65\degr \pm
2.5\degr$. The differences between the models are only marginally
visible on Fig.~\ref{f:seq}. However, comparing the models with the
top row of symmetrised data, it is noticeable that the velocity
dispersion is less well fitted with increasing inclination. On the
other hand, the fit to $h_{3}$ improves with higher
inclinations. Higher-order moments change similarly and the final
$\chi^{2}$ is the result of this combined effect. Surprisingly, the
difference between the best-fit model and the data are bigger than the
difference between the other models and the best-fit model, provoking
a question whether the determination of the inclination is robust.

Recently, EGF03\nocite{2003MNRAS.345.1297E} used two-integral
$f(E,L_z)$ dynamical models and found a best fit for an inclination of
60\degr~but also stated that models with $58\degr \le i \le 65\degr$
fit equally well. Our three-integral models have a larger freedom in
fitting the observations and it is unknown whether these models can
uniquely constrain the inclination. Our obtained inclination is close
to the previous measurements in the literature (references in
Sections~\ref{s:intro} and~\ref{s:obs}), and also to the inclination
measured from gas and dust. NGC~2974 is perhaps a special case (in
terms of its intrinsic structure and geometry) for which
three-integral models are able to give a stronger constraint on the
inclination. We return to this issue in Section~\ref{ss:param}.

The integral space of the best-fitting model, (i.e. the space defined
by the isolating integrals of motion ($E,L_{z},I_{3}$) that define the
orbits and the DF $f=f(E,L_{z},I_{3})$), is shown in
Fig.~\ref{f:int_space}. Each panel presents mass assigned to orbits of
constant energy, parameterised by the radius of the circular
orbit. This orbit also has the maximum angular momentum and circular
orbits of negative and positive angular momentum are in the bottom
left and right corner of each plot, while the low angular momentum
orbits are close to the symmetry ($y$) axis. An interesting feature
dominates the panels with radii larger then 5\farcs11. A high fraction
of mass is assigned to orbits with high angular momentum.  This
indicates that the bulk of the stars between these radii rotate with
high angular momentum.  A possible physical interpretation is that a
large fraction of the stars orbit in a
disc. \citet{1994MNRAS.270..325C} argued that NGC~2974 has an embedded
stellar disc and in their two-integral Jeans models they were not able
to fit the stellar kinematics without introducing a disc with
$\approx7\%$ of the total galaxy light.  On the other hand, the
two-integral models of EGF03\nocite{2003MNRAS.345.1297E} did not need
to invoke an additional stellar disc component to fit data consisting
of their {\tt TIGER} data and the three long slits of
\citet{1994MNRAS.270..325C}.  The integral space presented here
suggests that the three-integral models need orbits with high angular
momentum, but the selected orbits also have different values of
$I_{3}$, and therefore do not represent a very thin stellar disc as
assumed by \citet{1994MNRAS.270..325C}, but a somewhat flattened
distribution of stars like in a normal S0.  The relative light
contribution of the high-angular momentum orbits is $\approx10\%$,
which corresponds to a total stellar mass of $1.5\times10^{10}\Msun$
assuming the best-fit model inclination and $\Upsilon$.\looseness=-1

%
%

\section{Tests of Schwarzschild's orbit-superposition models}
\label{s:tests}
In the previous section we used three-integral models to recover the
inclination, mass-to-light ratio and the internal structure of
NGC~2974. Surprisingly, we find that the inclination is tightly
constrained by $\chi^2$ contours (Fig.~\ref{f:grid}), although the
difference between the models (Fig.~\ref{f:seq}) are smaller than the
difference between formally the best-fitting model and the data. In
this section, we wish to test the robustness of those results as well
as the general ability of the three-integral models to recover the
given parameters. For this purpose we constructed an axisymmetric
model mimicking NGC~2974 using two integrals of motion: the energy
$E$, and the $z$-component of the angular momentum, $L_{z}$. This
two-integral galaxy model has the advantage of a known DF,
$f=f(E,L_{z}$), everywhere, which we want to compare with the results
of the three-integral modelling. There are three issues we wish to
test:

\begin{itemize}

\item[1.] Recovery of the input parameters of the two-integral model
          galaxy. This is a general test to show whether the
          three-integral method can recover the parameters used in
          construction of a test model. We wish to be consistent with
          the observations and consider only the recovery of the input
          mass-to-light ratio and inclination, especially in light of
          the results from the Section~\ref{s:dyno}.

\item[2.] Influence of the spatial coverage. The \SAURON kinematic
          observations of NGC~2974 roughly cover one effective
          radius. This is also the typical size of most kinematic
          observations of other early-type galaxies from the
          \SAURON~sample. Here we want to test the influence of the
          limited extent of the kinematic coverage on the recovery of
          the orbital distribution. We do this by comparing the
          difference between models using a \emph{limited} and a
          \emph{full} spatial kinematic information provided by the
          two-integral galaxy model.

\item[3.] The recovery of the input DF. We wish to test the ability of
          our three-integral models to correctly recover the true
          input (two-integral) DF. Similar tests were also presented
          by \citet{2004MNRAS.353..391T} for their implementation of
          Schwarzschild's method.  \citet{1999ApJS..124..383C} and
          \citet{ver02} and \citet{2004MNRAS.347L..31C} described
          similar tests using two-integral Schwarzschild models.
\end{itemize}

\subsection{The input two-integral test model}
\label{ss:2I}
An $f(E,L_z)$ model of NGC~2974 was constructed using the
\citet{1993MNRAS.262..401H} contour integration method (hereafter HQ,
see also \citealt{1995MNRAS.274..602Q}). We used the mass model from
Section~\ref{ss:mass} parameterised by MGE. This approach follows in
detail \citet{1999MNRAS.303..495E}, and, due to the properties of
Gaussian functions, simplifies the numerical calculations
significantly. Given the density of the system, the HQ method gives
the unique even part of the DF,
$f_{e}=\frac{1}{2}[f(E,L_{z})+f(E,-L_{z})]$ (even in $L_{z}$). The odd
part can be calculated as a product of the even $f_{e}$ and a
prescribed function $h=h(L/L_{z})$. The magnitude of the function $h$
is chosen to be smaller than unity, which ensures that the final DF,
$f=f_{e}+f_{o}$, is physical (i.e. non-negative, provided that $f_{e}
>> 0$ everywhere). In practice, the odd part is chosen to fit the
observed kinematics (mean streaming) by flipping the direction of
orbits with respect to the symmetry axis (photometric minor axis).

The two-integral model of NGC~2974 was computed using $i=60\degr$ and
$\Upsilon=4.6$. In order to construct a realistic DF we also included
a black hole with mass $M_{BH} = 2.5\times10^{8} \Msun$, from the
$M_{\rm BH} - \sigma$ relation. The MGE model, constructed from a
finite spatial resolution HST WFPC2 image, has by construction an
unrealistic flat asymptotic density profile well inside the central
observed pixel ($r\la0\farcs02$) and therefore we assumed a cusp with
a power-law slope of 1.5 ($\rho = r^{-1.5}$) inside that radius,
following the prescription of \citet{1999MNRAS.303..495E}. The final
two-integral DF was computed on a fine adaptive grid (i.e. more points
in the region of strongly-changing DF with $E/E_{max} < 2$, where
$E_{max}$ is the value of the central potential of the model excluding
the black hole) of $140\times79$ points in ($E,L_{z}$). A fine grid of
LOSVDs was computed from this DF. These LOSVDs were used to compute
the observable LOSVDs on 3721 positions, two-dimensionally covering
one quadrant of the sky plane ($50\arcsec \times 50\arcsec$),
accounting for the instrumental set up (size of \SAURON pixels) and
atmospheric seeing (which matched the observations of NGC~2974 and was
used for the three-integral models in Section~\ref{ss:3I}). The
parameters of the two-integral model of NGC~2974 are listed in the
first column of Table~\ref{t:2I}.

\begin{figure}
\begin{center}
  \includegraphics[width=\columnwidth]{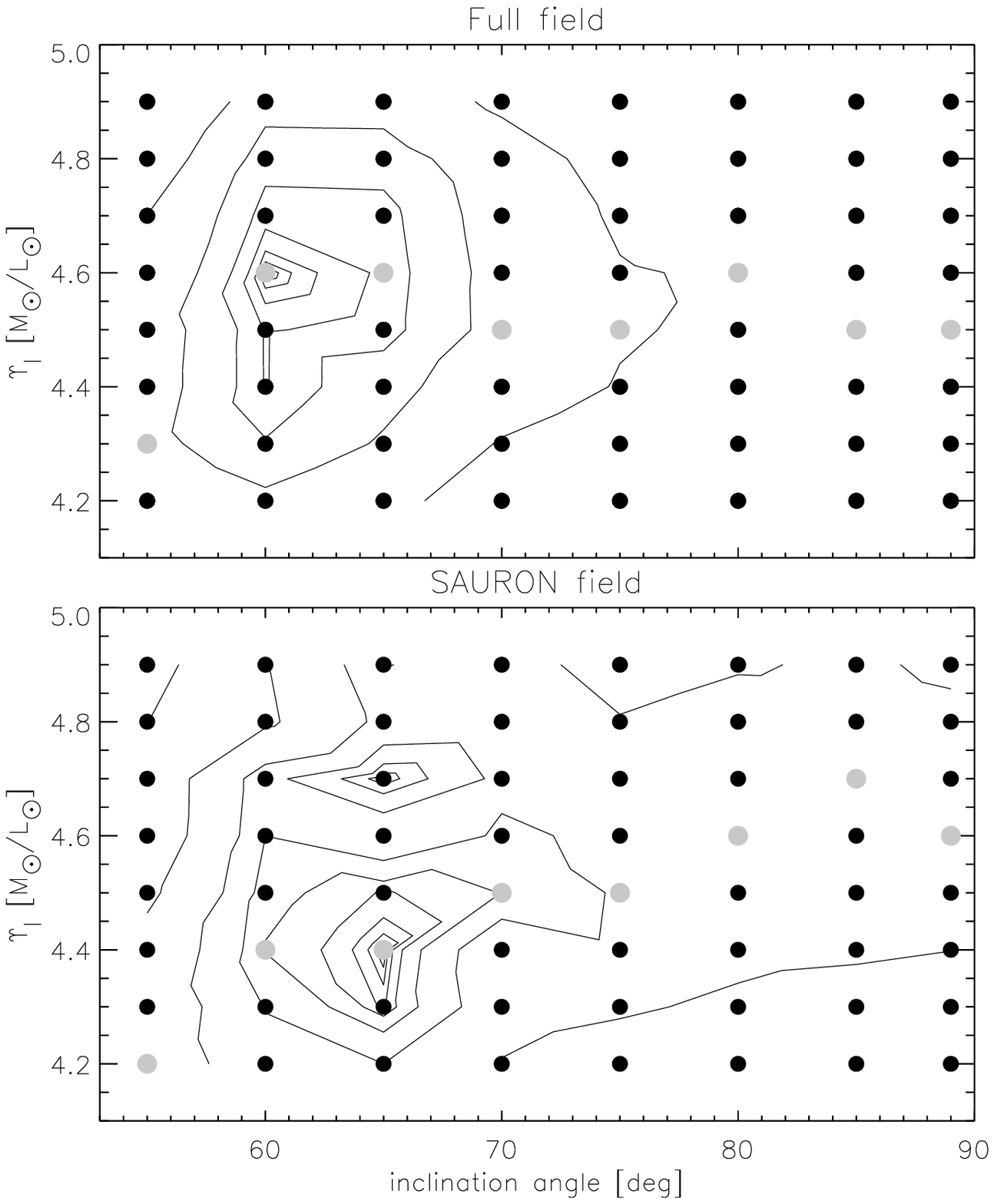}
  \caption{\label{f:2Igrids} Three-integral model grids of inclination
  angle versus $M/L$ ratio, $\Upsilon$. Contours present constant
  $\Delta \chi^{2}$, measuring the goodness of fit of the
  models. Every symbol corresponds to a three-integral axisymmetric
  model with given inclination angle and $\Upsilon$. The grey symbols
  indicate the best fitting models at a given inclination. The top
  grid presents models using the \emph{Full field} set of kinematic
  constraints. The global minimum is for the model with $i=60\degr$
  and $\Upsilon=4.6$. The bottom panel presents models using the
  \emph{\SAURON field} set of kinematic constraints. The global
  minimum here is for the model with $i=65\degr$ and $\Upsilon=4.4$. }
\end{center}
\end{figure}

\begin{table}
   \caption{The properties of two-integral models and comparison to
   three-integral best-fit results.}
   \label{t:2I}
$$
   \begin{array}{c|ccc}
       \hline
       \noalign{\smallskip}
       &$2I model$ & $Full field$  & $\SAURON field$\\
       & (1)&(2)&(3) \\
       \noalign{\smallskip}
       \hline
        \Upsilon   & 4.6     & 4.6  \pm 0.1   & 4.4 \pm 0.1\\
         i       & 60\degr & 60\degr \pm 5\degr& 65\degr \pm 5\degr\\
       \noalign{\smallskip}
       \hline
   \end{array}
$$ {Notes -- Col.(1): parameters of the two-integral model of
NGC~2974; Col.(2): recovered parameters using the \emph{Full field}
spatial coverage ($100\arcsec \times 100\arcsec$ effectively) of
kinematic constraints; Col.(3): recovered parameters using the
\emph{\SAURON field} spatial coverage ($40\arcsec \times 40\arcsec$
effectively) of kinematic constraints.}
\end{table}

\begin{figure*}
\begin{center}
  \includegraphics[width=\textwidth]{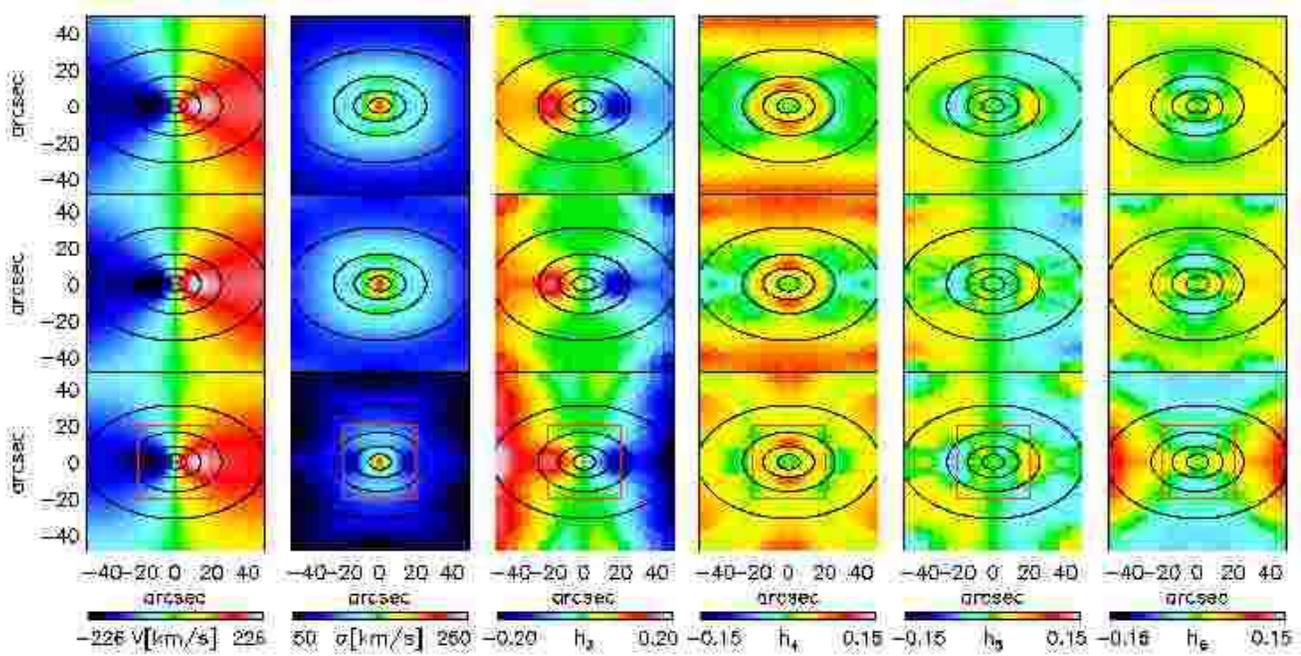}
  \caption{\label{f:2Imod} Comparison between two-integral analytical
  model kinematics and three-integral models. First row: two-integral
  model used as input to the three-integral code. Second row: best
  fitting three-integral model ($i=60\degr,\Upsilon=4.6$) using the
  \emph{Full field} set of kinematic constraints. Third row:
  best-fitting three-integral model ($i=65\degr,\Upsilon=4.4$) using
  the \emph{\SAURON field} set of kinematics. The spatial extent of
  this set is marked by red squares on the maps. From left to right
  each panel presents: mean velocity $V$, velocity dispersion
  $\sigma$, and Gauss-Hermite moments: $h_{3}, h_{4}, h_{5}$ and
  $h_{6}$. Isophotal contours of total light are shown with solid
  lines.}
\end{center}
\end{figure*}

In order to mimic the real observations (as well as to reduce the
number of observables in the fit) we adaptively binned the spatial
apertures using the Voronoi tessellation method of
\citet{2003MNRAS.342..345C} as it was done for the observations,
assuming Poissonian noise. The final LOSVDs were used to calculate the
kinematic moments (V, $\sigma$, $h_{3}$ to $h_{6}$) by fitting a
Gauss-Hermite series (first row on Fig.~\ref{f:2Imod}). These values
were adopted as kinematic observables for the three-integral
models. From these data we selected two sets of kinematic observables:

\begin{itemize}

\item The first set consisted of all 513 spatial bins provided by the
      two-integral model. In terms of radial coverage this \emph{Full
      field} set extended somewhat beyond two effective radii for NGC~2974.

\item The second set had a limited spatial coverage. It was limited by
      the extent of the \SAURON observations of NGC~2974,
      approximately covering one effective radius. We call this
      observational set of 313 bins the \emph{\SAURON field}.

\end{itemize}

\subsection{Recovery of input parameters}
\label{ss:param}
Our two-integral models are axisymmetric by construction and all
necessary information is given in one quadrant bounded by the symmetry
axes (major and minor photometric axes). Hence, the inputs to the
three-integral code covered only one quadrant of the galaxy. Although
only one quadrant was used for the calculations we show all maps
unfolded for presentation purposes.

The three-integral models were constructed in the same way as
described in Section~\ref{ss:3I}. For both (\emph{Full field} and
\emph{\SAURON field}) sets of kinematic data we created orbit
libraries of $2\times 41\times10\times10$ orbits and constructed
models on grids of ($\Upsilon,i$). As for the real data, we used a
regularisation parameter $\Delta=4$. The resulting grids are presented
in Fig.~\ref{f:2Igrids}, and the best-fitting models are listed in the
Table~\ref{t:2I}. The three-integral models were able to recover the
true input parameters within the estimated errors, although the
\emph{\SAURON field} models were less accurate.

\begin{figure*}
\begin{center}
  \includegraphics[width=\textwidth]{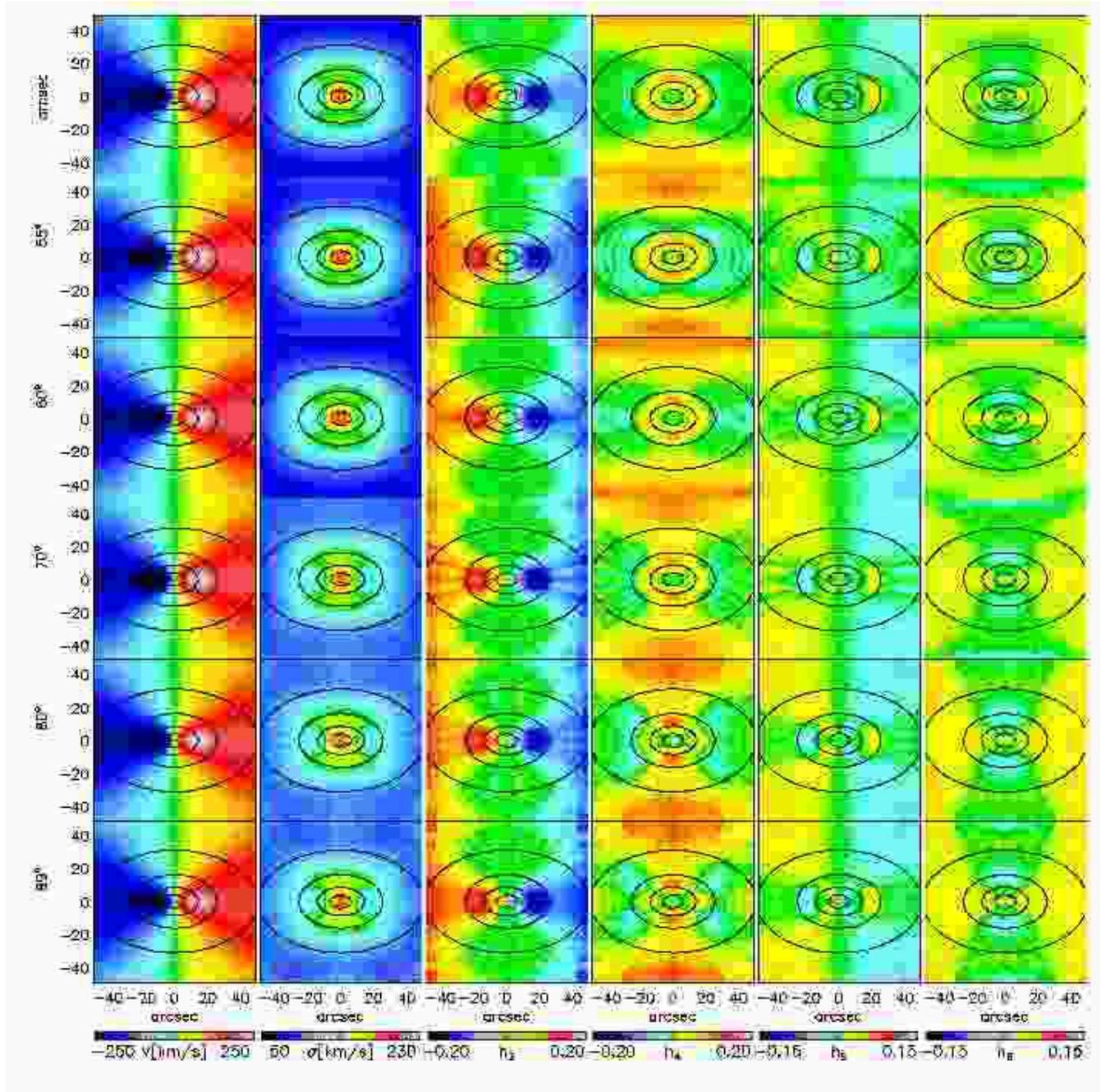}
  \caption{\label{f:2Iseq} Sequence of three-integral models for
  different inclinations fitting the (\emph{Full field} of kinematic
  observables. First row: two-integral test model. Subsequent rows:
  three-integral models for $i=(55\degr, 60\degr, 70\degr, 80\degr,
  89\degr)$. Columns from left to right present moments of the LOSVD:
  $v$, $\sigma$, and from $h_{3}$ to $h_{6}$. Isophotal contours of
  total light are shown with solid lines.}
\end{center}
\end{figure*}

The kinematic observables computed from the two-integral models are
noiseless, without errors nor intrinsic scatter typical of real
measurements. For each kinematic observable we assigned a constant
error, but representative to the \SAURON observations of NGC~2974:
$1\sigma$ errors for $V$, $\sigma$, $h_{3}$, and $h_4 \ldots h_{6}$
were 4$\kms$, 7$\kms$, 0.03 and 0.04 respectively.

As we wanted to test an ideal situation we computed the kinematic
observables from the two-integral model without adding noise
(intrinsic scatter) typical for real measurements. However, given the
fact that the input model is noiseless, the $\chi^2$ levels computed
from the fit to the kinematics are meaningless. In order to have an
estimate of the uncertainties in the recovery of the parameters we
computed half a dozen Monte Carlo realisations of the kinematic data,
introducing the intrinsic scatter to the noiseless data. For each
Monte Carlo data sets we constructed a three-integral model grid like
in Fig.~\ref{f:2Igrids}. Due to time limitations we calculated
parameter grids ($\Upsilon$, $i$) of models with smaller orbit
libraries ($2\times21\times7\times7$ orbits) of both \emph{Full field}
and \emph{\SAURON field} data sets. In this case we also applied the
regularisation scheme with $\Delta=4$. Approximate $3\sigma$
confidence levels assigned to the best-fitting parameters are listed
in Table~\ref{t:2I}. Setting the regularisation to zero, we observed
similar trends.

The numbers in Table~\ref{t:2I} suggest that the inclination is
formally recovered by three-integral models, as seen in the case of
the observation (Section~\ref{ss:model}). We repeat the exercises of
plotting a sequence of models with different inclinations (using the
\emph{Full field} kinematics). They are presented in
Fig.~\ref{f:2Iseq} and we can see a very similar trend as in
Fig.~\ref{f:seq}: there appears to be little difference between the
models, although by scrutinising the details, it is possible to choose
the best model by eye.

\begin{figure*} \centering
{\vbox { \epsfxsize=\textwidth
\epsfbox{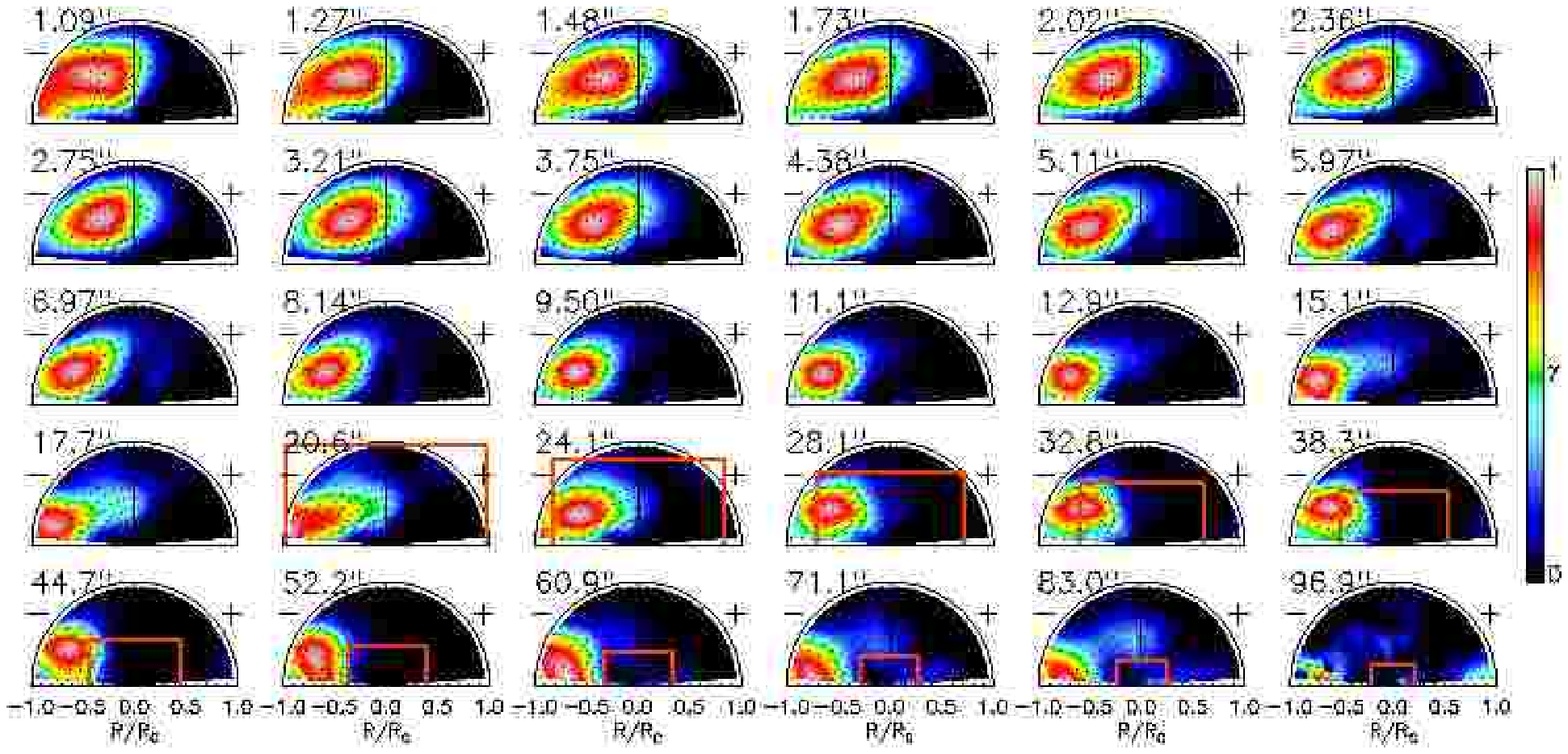}
\epsfxsize=\textwidth
\epsfbox{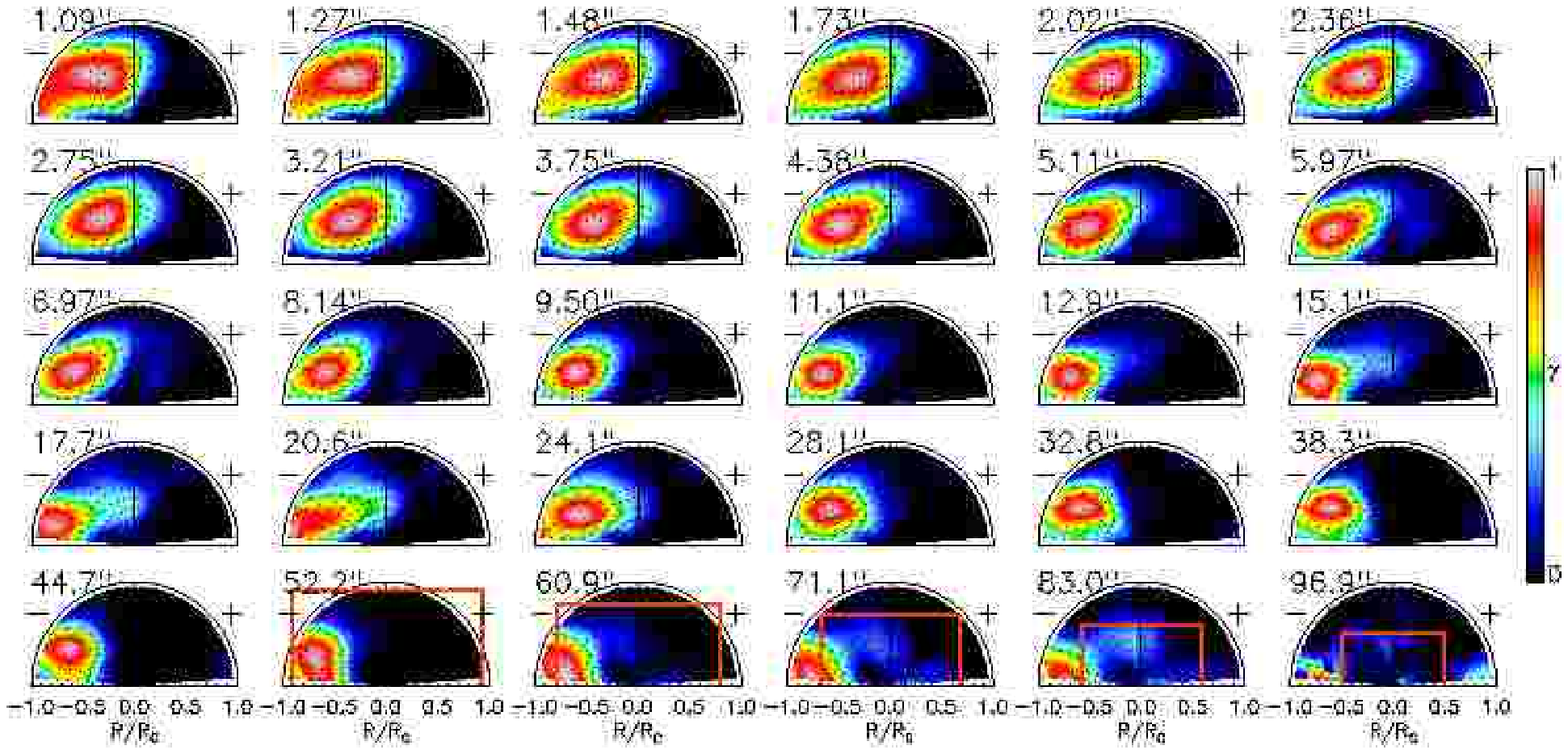} }}
\caption{\label{f:2I_int_space} Comparison of the integral spaces of
        the $f(E,L_z)$ test models. Upper five rows belong to the
        model constrained by the \emph{\SAURON field} kinematics and
        the lower five rows to the model constrained by the \emph{Full
        field} kinematics. The meaning of each panel is the same as in
        Fig.~\ref{f:int_space}. The colour bar on the right represents
        the relative colour coding of the orbital mass weights
        ($\gamma$), where 1 represent the largest mass weight assigned
        to an orbit at the given energy. In the region constrained by
        both kinematic sets the integral spaces are
        indistinguishable.}
\end{figure*}

The smoothness of the data of the two-integral test model helps in
recognising the best-fit model. The models with lower inclination
(towards face-on) are generally smoother than the higher inclination
(towards edge-on) models, which systematically show radial
structures. These ``rays'' visible on Fig.~\ref{f:2Iseq} are artifacts
of the discreteness of our orbit library. The starting points of the
orbits correspond to the positions of orbital cusps, which carry the
biggest contribution to the observables. The total number of cusps is
determined by the number of orbits, and the finiteness of the orbit
library is reflected in the discrete contributions of the cusps to the
reconstructed observables (see Fig.~1.2 of
\citealt{2004cbhg.sympE...5C}).  The projection effects, however,
increase the smoothness by spatially overlapping different cusps;
hence, models projected at e.g. $i=60\degr$ will be smoother than
models viewed edge on. We believe this effect could influence the
$\chi^{2}$, favouring the lower inclination models.

Although present, this effect does not provide the only constraint on
the inclination. When examined more closely, models with low
inclination do reproduce certain features better, for example: the
shape and amplitude of the velocity dispersion in the central
20\arcsec, $h_{3}$ and $h_{4}$ at the larger radii (towards the edge
of the field), the central 10\arcsec\ of $h_{5}$ and $h_{6}$. In all
cases the model with $i=60\degr$ reproduces these features better than
other models.  Comparing the models, the most significant contribution
to the $\chi^{2}$ comes from the velocity dispersion, but generally,
individual observables have slightly different $\chi^{2}$ values,
which increase with the inclination moving away from 60\degr~and is
visible only as a cumulative effect. This explains the similarities of
the different models to the eye, although they are formally
significantly different.

The difference in the model observables (which include moments up to
$h_{6}$) are below the level of the expected systematics in the data
(e.g. template mismatch), or in the models (e.g. regularisation or
variations in the sampling of observables with orbits). In the case of
NGC~2974, using high signal-to-noise two-dimensional data, the
difference between the models themselves are smaller than between the
best-fitting model and the data, implying that the inclination is only
weakly constrained. This result suggests a fundamental degeneracy for
the determination of inclination with three-integral models, which is
contrary to indications from previous work by
\citet{2002MNRAS.335..517V}. Theoretical work and more general tests
on other galaxies are needed for a better understanding of this issue.

\subsection{Effect of the field coverage on orbital distribution}
\label{ss:phcomp}
The next step is to compare in more detail the three-integral models
using the two different kinematic data sets. The kinematic structures
of the best-fitting models are presented in Fig.~\ref{f:2Imod}. In the
region constrained by the kinematic data both models reproduce equally
well the input kinematics. As expected the regions outside the
\emph{\SAURON field} are not reproduced well. It is more interesting
to compare the phase spaces of the models.  In particular, we wish to
see whether the mass weights assigned to the orbits (represented by
the integrals of motion) are the same for the two models.

The corresponding integral-spaces are shown in Fig.~\ref{f:DFcomp}.
Red rectangular boxes represent the extent of the kinematics used to
constrain the models. In the regions constrained by both kinematic
sets, the two integral spaces are identical: both models recover the
indistinguishable orbital mass weights. The differences appear at
larger radii (beyond 20\arcsec), outside the area constrained by the
\emph{\SAURON field} kinematic set. Putting this result in the
perspective of observations, the resulting phase space (in the region
constrained by the observations) does not depend on the extent of the
radial coverage used to constrain the model. This result strengthens
the case of the NGC~2974 modelling results, where we have
integral-field observations reaching $\approx 1$ r$_{e}$. The
recovered integral space and its features would not change
significantly if we had a spatially larger observational field.

\subsection{Recovery of the internal moments}
\label{ss:internal}
\begin{figure}
\begin{center}
  \includegraphics[width=\columnwidth]{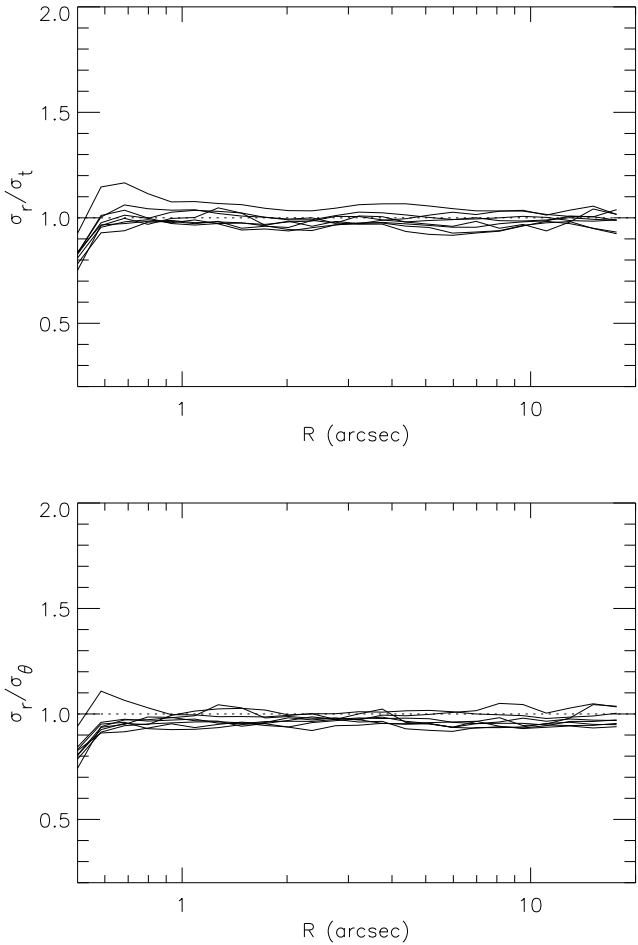}
  \caption{\label{f:2Imoments} Moments of the velocity ellipsoid for
  the two-integral test galaxy recovered by the three-integral model
  (constrained by the \emph{Full field} kinematic set). The upper
  panel shows the ratio of the radial and tangential moment as defined
  in the text. The lower panel presents the ratio of the radial and
  latitudinal moments. Different lines show the ratio of the moments
  in the meridional plane at different position angles, starting from
  the major-axis to the minor axis. In a true two-integral galaxy, all
  three moments of the velocity ellipsoid have to be equal. Note that
  the deviation inside 1\arcsec are expected since the data do not
  constrain the model in that region. }
\end{center}
\end{figure}

We wish to see if the best-fitting three-integral model to the
$f(E,L_z)$ test galaxy model is consistent with the input, i.e.,
whether three-integral models will recognise the true structure of the
test galaxy. A first estimate can be achieved by investigating the
internal structure of the resulting model galaxy, specifically the
shape of the velocity ellipsoid. We define the tangential dispersion
as $\sigma_{t} = [\frac{1}{2}(\sigma^{2}_{\theta} +
\sigma^{2}_{\phi})]^{1/2}$. Note that $\sigma_{\phi}$ includes only
random motion so that for an isotropic distribution, under the given
definition, the radial ($\sigma_{r}$) and tangential dispersion are
equal. Since two-integral models are isotropic in the meridional plane
per definition, we expect to recover that $\sigma_{r}$ is equal to
$\sigma_{\theta}$ and the cross-term $\sigma_{R \theta}$ is equal to
zero. In Fig.~\ref{f:2Imoments} we show the ratio of the moments of
the velocity ellipsoids at different positions in the meridional plane
and at different radii from the model constrained by the \emph{Full
field} kinematics. One can see that $\sigma_{r} = \sigma_{\theta}$
within $\approx5\%$, confirming that our three-integral model recovers
the true internal moments. Also, computing the cross-terms $\sigma_{R
\theta}$, we verified that it is negligible everywhere. Similar
results are also recovered from the \emph{\SAURON field} model: in the
constrained region the ratio of $\sigma_{r}$ and $\sigma_{\theta}$
moments are consistent with unity.

\subsection{Recovery of the distribution function}
\label{ss:rdf}

\begin{figure}
\begin{center}
  \includegraphics[width=\columnwidth]{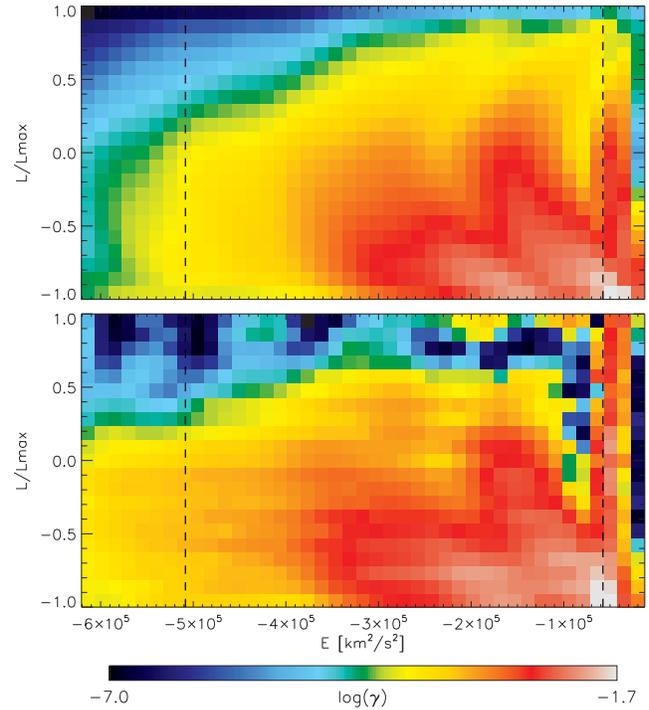}
  \caption{\label{f:DFcomp} Comparison of the mass weights from the
  two-integral model (upper panel) and the results of the
  three-integral modelling (lower panel) using the two-integral model
  as input. The galaxy centre is on the left side. The first bin
  containing the contribution of the black hole was not plotted, since
  the resolution of the models do not allow for its recovery. The two
  vertical lines enclose the region constrained by the kinematic
  data. The colour bar represents the colour coding of the orbital
  mass weights ($\gamma$) in log scale assigned to any interval of
  ($E,L_z$). There are two main differences between the input and the
  best fit model: (i) the horizontal stripes, which are due to the
  discrete way in which we select orbits in $L_z$ range, and (ii) the
  mass weights at high $L_z$ values, which, however, contain a very
  small fraction of the galaxy mass and are not significant.}
\end{center}
\end{figure}

The previous result shows that the constructed three-integral model is
a consistent representation of the input $f(E,L_z)$ model. A more
conclusive test, however, is to compare the distribution functions.

The results of the three-integral Schwarzschild method are orbital
mass weights, $\gamma'_{3I}(E,L_{z},I_{3})$, for each set of integrals
of motion ($E$, $L_{z}$, $I_{3}$), which define an orbit. The DF is
related to the mass weights via the phase-space volume (for a detailed
treatment see \citealt{1984ApJ...287..475V}):
\begin{eqnarray} \label{eq:DFmw}
\lefteqn{\gamma (E,L_{z}, I_{3})\; \ud E \ud L_{z} \ud I_{3} = {}} \nonumber \\
& & f(E,L_{z},I_{3})\; \Delta V(E,L_{z},I_{3})\; \ud E \ud L_{z} \ud I_{3},
\end{eqnarray}
where
\begin{equation} \label{eq:deltaV}
\Delta V(E,L_{z},I_{3}) = \int \limits_{\Omega} \mathcal{J}(\vec{x},E,L_{z},
I_{3}) \ud^{3}x,
\end{equation}
and $\mathcal{J}(\vec{x},E,L_{z},I_{3})$ is the Jacobian of the
coordinate transformation from $(\vec{x}, \vec{v})$ to
$(\vec{x},E,L_{z},I_{3})$, and $\Omega$ is the configuration space
accessible to an orbit defined by the integrals ($E$, $L_{z}$,
$I_{3}$). Unfortunately, $I_{3}$ is not known analytically, so the
above relation can not be explicitly evaluated, except for separable
models. For this reason we limit ourselves to test the
\emph{consistency} of our three-integral mass weights with the input
two-integral DF. This is possible, since if the recovered
three-integral DF is equal to the input DF, then the mass
$\gamma_{2I}(E,L_{z})$ assigned to the stars in a given range of
$(E,L_{z})$ by the three-integral model has to be equal to the mass in
the same range of the input model.

There exists a precise relation between the input two-integral DF,
$f=f(E, L_{z})$, and the corresponding orbital mass weights,
$\gamma_{2I}(E,L{z})$. The total mass of the system is the integral of
the DF over the phase-space. Using this, \citet{1999ApJS..124..383C}
derived the expression for the mass weights in Appendix B of their
paper (eq. B4):
\begin{equation} \label{eq:2Iinta}
 \gamma_{2I}(E,L_{z}) = \int\limits_{\Omega} \frac{\ud M}{\ud E \ud
L_{z}}\ud E \ud L_{z} \\
\end{equation}
with
\begin{equation}\label{eq:2Iintb}
 \frac{\ud M}{\ud E \ud L_{z}} = f(E,L_{z}) \times
 \oint\limits_{ZVC(E,L_{z})} (R \ud z - z \ud R)
\end{equation}
where the contour integral yields the area of the zero velocity curve
(ZVC). Before applying eqs.~(\ref{eq:2Iinta}) and~(\ref{eq:2Iintb}) on
the two-integral DF we rebinned it to the same grid of $\Delta E$ and
$\Delta L_{z}$ as the three-integral mass weights. Finally, we
approximate the integral~(\ref{eq:2Iinta}) by multiplying the mass
fraction in each grid cell by $\Delta E \Delta L_{z}$.

For our comparison we defined as the energy intervals the set of
energies used in the construction of the three-integral models (total
of 41). The interval in angular momentum was defined as a step of 0.1
of $L_{z}/L^{max}_{z}$ from $-1$ to $+1$ (total of 20) for a given
energy. The resulting grid of orbital weights is relatively coarse,
but is representative of the model.  The agreement between the two
sets of mass weights is shown in Fig.~\ref{f:DFcomp}. The main
features of the given two-integral test model are well reproduced by
the three-integral model. Again, the mass weights should be compared
in the region constrained by the data (between the vertical lines in
the figure).  In order to quantify the agreement we restricted the
comparison to the part of the distribution of mass weights
contributing significantly to the model inside this region, roughly
bounded by the green levels on Fig.~\ref{f:DFcomp}. This selected
region contains about 97\% of mass.  The mean absolute deviation
between the two sets of mass weights is 6\%, with peaks around 15\%,
except for a narrow region towards the edge of the kinematic
coverage around $E= - 8 \times 10^4 km^{2} s^{-2}$ and with
$L_{z}/L^{max}_{Lz} > 0$, where deviations can reach 35\%. The mass
weights of the three-integral model are also relatively noisy which is
mostly the consequence of imposed discreteness as well as the
numerical nature of the method.

From our test model, where the potential is known, we conclude that
the Schwarzschild method can reliably recover the input distribution
function within the region constrained by two-dimensional kinematics
(including higher-order moments). We have already shown that the
existence of constraints outside of an effective radius does not
significantly affect the orbital distribution {\it inside} an
effective radius. This suggests that, in the region constrained by
integral-field kinematics, a representative DF can be recovered using
the Schwarzschild method.

%
%

\section{Modelling of emission-line gas}
\label{s:em_gas}
Clearly, with its prominent gas component, NGC~2974 is an unusual
elliptical galaxy. The observations indicate the morphological
similarity between the small (H{\small II}) and large scale (H{\small
I}) gas discs. Also, long-slit measurements of stellar and gas motions
detect similarities between the stellar and gas kinematics
\citep{1989ApJ...346..653K,1994MNRAS.270..325C}. In this section we
investigate the inclination of the gaseous component as well as
construct dynamical Jeans model of the gas disc as is
\citet{1994MNRAS.270..325C}, in order to compare them to the results
of the stellar dynamical modelling.

\subsection{Inclination of the gas disc}
\label{ss:incl}
The existence of the emission-line gas disc in NGC~2974 can be used to
infer the inclination of the galaxy assuming an equilibrium dynamical
configuration. This has been attempted before and in all studies the
inclination of the gas disc was consistent with $55\degr - 60\degr$
(\citealt{1993sdce.conf..225A} 55\degr; \citealt{1993A&A...280..409B}
59\degr; CvdM94\nocite{1994MNRAS.270..325C} 57.5\degr;
\citealt{1998A&AS..128...75P} 60\degr). The high quality of the
two-dimensional \SAURON kinematics allows us to estimate the
inclination of the emission-line gas disc more accurately.

\begin{figure}
        \includegraphics[width=\columnwidth]{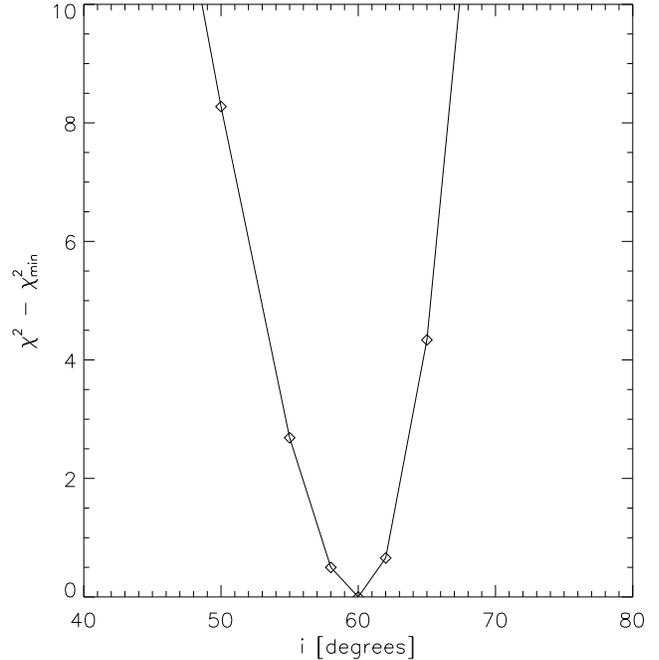}
        \caption{\label{f:chi_incl} $\Delta\chi^{2}$ as a function of
        inclination obtained by comparing the symmetrised data for
        NGC~2974 and the model gas disc velocity map described
        in~\ref{ss:incl}. }
\end{figure}

\begin{figure*}
  \hspace{-1.5cm}
        \includegraphics[width=0.9\textwidth]{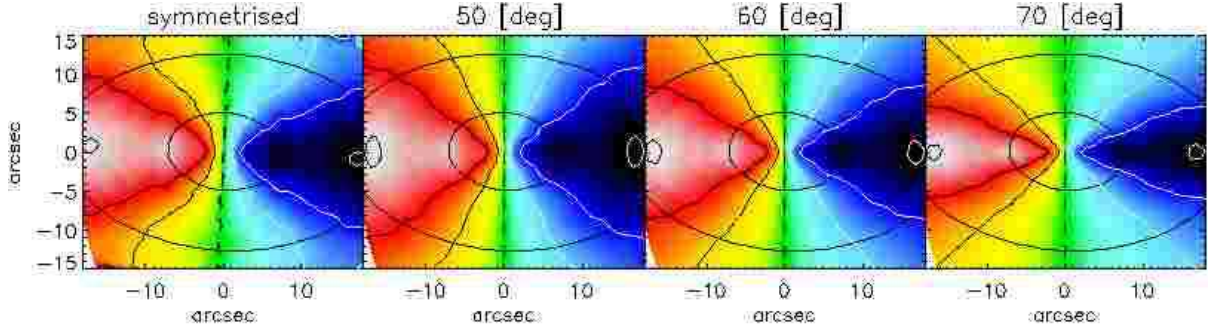}
        \caption{\label{f:incl_com} Comparisons between observed and
        model gas disc velocity fields. From left to right:
        observations (symmetrised), model fields for inclinations of
        50$\degr$, 60$\degr$ (best-fitting), 70$\degr$. Isophotal
        contours of total light are presented with ellipses. }
\end{figure*}

We assume the motion of the emission-line gas is confined to a thin
axisymmetric disc. We neglect the deviations from axisymmetry
discussed in Section~\ref{s:kin}, and symmetrise the gas velocity map
in the same way as the stellar velocity map in
Section~\ref{ss:model}. This velocity map is an axisymmetric
representation of the observed field, which can be compared to an
axisymmetric model of the disc velocity maps. Constructing a model
axisymmetric two-dimensional velocity map requires only the kinematic
major axis velocity profile $v'_{mj}$. The entire map is then given by
the standard projection formula:
\begin{equation} \label{eq:vproj}
  v_{LOS}(x',y') = v_{\phi} \Big(\frac{x' \sin i}{r} \Big) = v'_{mj}
  \Big(\frac{x'}{r} \Big),
\end{equation}
where $r^{2} = x'^{2} + (y'/\cos i)^{2}$ and $i$ is the inclination of
the disc. It is clear from this formula that for a given observed
major-axis velocity, the velocity field is just a function of
inclination. From the first three odd coefficients ($c_{1}$, $c_{3}$
and $c_{5}$) of the kinemetric expansion we construct the velocity
profile $v'_{mj}$ along the major axis ($\theta$ =
$\overline{\phi_{1}}$). Using the major axis velocity profile, we
created a set of disc velocity fields inclined at different values of
$i$, and compared them with the symmetrised velocity field. We did not
correct for the influence of the PSF as this effect is small and is
confined to the central few arcseconds, which we excluded from the
comparison. We also compared the models with the non-symmetrised
velocity map and the results were in very good agreement, but with a
slightly larger uncertainty range.

Figure~\ref{f:chi_incl} presents the $\Delta\chi^{2}$ obtained by
subtracting the disc model velocity field from the symmetrised
measurements. The best-fitting inclination is $i = 60\degr \pm 3\degr$
(at $1\sigma$ level). Fig.~\ref{f:incl_com} shows a comparison between
the symmetrised and model velocity fields for a few representative
inclinations. The differences between the model fields are mostly in
the opening angle of the iso-velocity contours, which change with the
inclination of the field. The opening angle of the model with
$i=60\degr$ is the most similar the observed velocity field and this
significantly lowers the $\chi^{2}$ of the fit.

The best-fit inclination of $60\degr$ for the emission-line gas is in
excellent agreement with literature values determined from the various
gas components. Our best-fitting three-integral stellar dynamical
model was obtained for an inclination of $65\degr$ with $3\sigma$
uncertainty of $2.5\degr$. This inclination is close to the
inclination of $60\degr$ presented in this section, suggesting a good
agreement between the stellar and gaseous models. There is, however,
the concern that the agreement may not be as significant as it seems
in light of the tests and results from Section~\ref{ss:param}.

\subsection{A simple dynamical model for the disc}
\label{ss:asym}
At large scale, the gas kinematic maps are consistent with the
assumption that the emission-line gas is moving in a thin disc. This
assumption clearly breaks down in the inner few arcseconds, but at
this point we neglect this effect. The observed gas velocity
dispersion is high everywhere in the disc and is much larger than the
thermal velocity dispersion which should be of the order of
$\sigma_{thermal} \sim10 \kms$ \citep{1989agna.book.....O} and in any
case $< 30 \kms$. Clearly, in addition to the thermal dispersion, the
gas has another source of motion, which is not presently understood,
but is seen in many galaxies \citep{1995ApJ...448L..13B}. Several
studies \citep[e.g.][]{1998AJ....116.2220V, 2000AJ....120.1221V}
assumed that the non-thermal gas velocity dispersion is the result of
`local turbulence', without describing the details of the underlying
physical processes. In this assumption, the gas still rotates at the
circular velocity and the invoked turbulence does not disturb the bulk
flow of the gas on circular orbits. The alternative to this assumption
is that the non-thermal velocity dispersion component comes from
collisionless gravitational motion of the gas, where the gas acts like
stars: clumps of gas move on self-intersecting orbits. This is,
perhaps, not very physical for gas in general, but it can be applied
to estimate the difference between the circular and streaming velocity
(and including the projection effect on the observed velocity) of the
gas. Several studies used this approach successfully
\citep[e.g.][]{1994MNRAS.270..325C, 2000ApJ...536..319C,
2001ApJ...555..685B, 2003MNRAS.338..465A,
2004ApJ...605..714D}. Presently, the role and importance of the
asymmetric drift remains an unresolved issue.


\begin{figure}
        \includegraphics[width=\columnwidth]{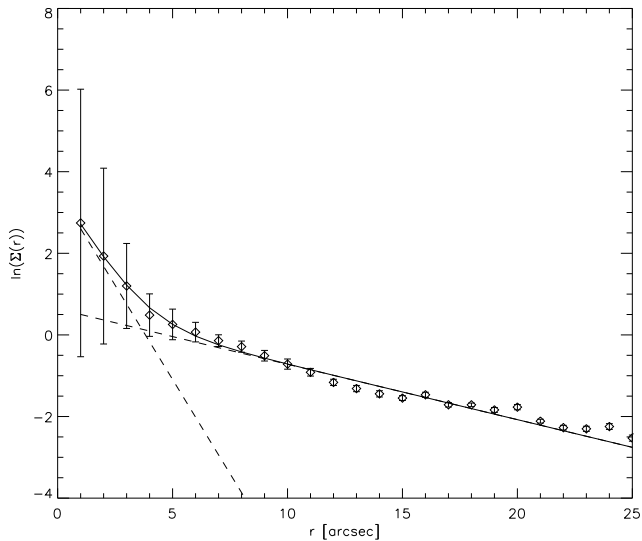}
        \caption{\label{f:dens} Fit to the surface-density profile of
        the [O{\small III}] emission lines. Dashed lines present
        individual exponentials given by eq.(\ref{eq:ro}). }
\end{figure}


\begin{figure}
        \includegraphics[width=\columnwidth]{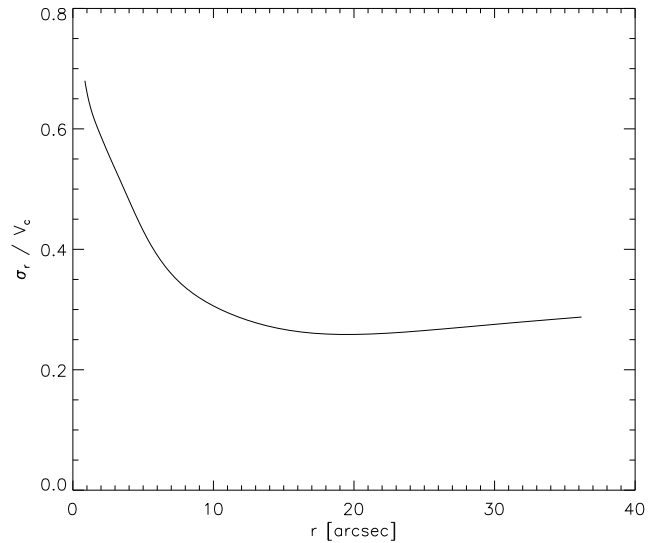}
        \caption{\label{f:asym_cond} Ratio $\sigma_{R}/V_{C}$ for the
        best-fitting model to the emission-line gas data.}
\end{figure}

In constructing our simple disc model we assumed that the
emission-line gas is moving in individual clumps that interact only
collisionlessly. The clumps move along ballistic trajectories in a
thin disc, under the influence of the galaxy potential given by the
stellar distribution (Section~\ref{s:dyno}). The gas kinematics are
determined by solving the Jeans equations for radial hydrostatic
equilibrium. Following \citet[][eq. 4-33]{1987gady.book.....B}, the
streaming velocity of gas can be written in cylindrical coordinates
as:
\begin{equation} \label{eq:v_fi}
  \bar{v}_{\phi}^{2} = V_{c}^{2} - \sigma_{R}^{2} \Big[ -R\frac{\ud\ln
  \rho}{\ud R} -R\frac{\ud\ln \sigma_{R}^{2}}{\ud R} - (1 -
  \frac{\sigma_{\phi}^{2}}{\sigma_{R}^{2}}) \Big]
\end{equation}
where we have assumed the distribution function depends on the two
classical integrals of motion, $f=f(E,L_{z})$, which implies
$\sigma_{R}$ = $\sigma_{z}$ and $\overline{v_{R}v_{z}}$ = 0. In
eq.~(\ref{eq:v_fi}), V$_{c}$ is the circular velocity ($\sqrt{R (\ud
\Phi/ \ud R)}$), $\Phi$ is the total potential of the galaxy obtained
from the MGE fit assuming inclination $i$, $\sigma_{R}$ and
$\sigma_{\phi}$ are the radial and azimuthal velocity dispersions, and
$\rho$(R) is the spatial number density of gas clouds in the
disc. Lacking any alternative, we use the surface brightness of the
gas to estimate $\rho$(R). Instead of using the actual measured values
of the [O{\small III}] and H$\beta$ flux, we parametrise the
emission-line surface brightness with a double exponential law,
\begin{equation} \label{eq:ro}
  \rho = \rho_{0} e^{-\frac{R}{R_{0}}} + \rho_{1} e^{-\frac{R}{R_{1}}},
\end{equation}
in order to decrease the noise. The parameters are obtained from
the fit to the [O{\small III}] data shown in (Fig.~\ref{f:dens}),
where the two-dimensional surface brightness was collapsed to a
profile by averaging along ellipses of constant ellipticity
(ellipticity of the galaxy) and position angle (PA of the galaxy). The
errors are standard deviations of the measurements along each ellipse.

The relation between radial and azimuthal velocity dispersions can be
obtained using the epicyclic approximation. This gives (following
eq. 4-52 in \citealt{1987gady.book.....B}):
\begin{equation} \label{eq:ep}
  \frac{\sigma_{\phi}^{2}}{\sigma_{R}^{2}} = \frac{1}{2} \Big( 1 +
  \frac{\ud \ln V_{c}}{\ud \ln R} \Big)
\end{equation}
This approximation is valid for small values of the
asymmetric drift $v_{\phi} - V_{c}$, or, in other words, in the limit
of a cold disc with small velocity dispersion, $\sigma \ll
V_{c}$. This is marginally the case in NGC~2974, clearly violated in
the central $<5\arcsec$, but it is acceptable in most of the observed
regions (Fig.~\ref{f:asym_cond}).

The observed quantities can be obtained from the calculated intrinsic
properties projecting at an inclination angle $i$. The projected
two-dimensional line-of-sight (LOS) velocity field is given by
eq.~(\ref{eq:vproj}). Within these assumptions of the disc model, the
projected LOS velocity dispersion is:
\begin{equation} \label{eq:slos}
  \sigma^{2}_{LOS} = (\sigma_{\phi}^{2} - \sigma_{R}^{2})
  \Big(\frac{x' \sin i}{r} \Big)^{2} + \sigma_{R}^{2}
\end{equation}

\begin{figure}
        \includegraphics[width=\columnwidth]{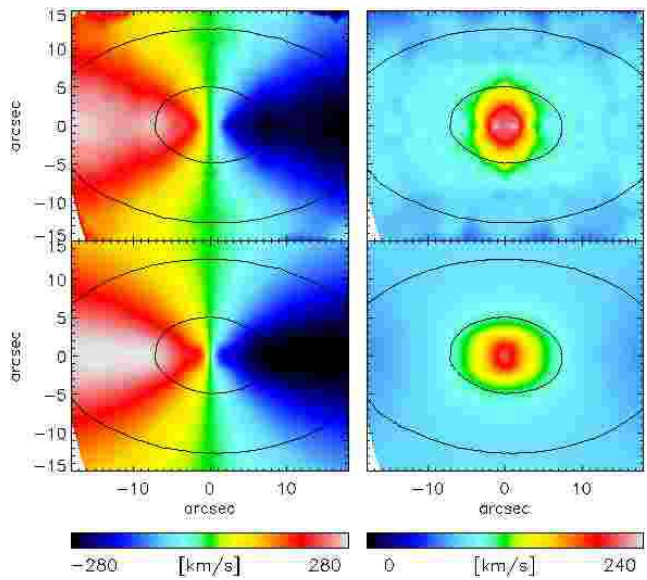}
        \caption{\label{f:asym}\emph{Bottom Panels:} Asymmetric drift
        model for the best fitting parameters compared to \emph{Top
        Panels:} the symmetrised (mirror-(anti)-symmetric filtering
        with 6 terms) mean velocity (first column) and velocity
        dispersion (second column). Overplotted isophotes are levels
        of the reconstructed total intensity.}
\end{figure}

We constructed the asymmetric drift models of the emission-line gas in
NGC~2974 using the MGE parametrisation of the potential,
$\Upsilon$=4.5, inclination $i=60\degr$, and simultaneously accounting
for the atmospheric seeing and pixel size of the \SAURON observations
(see \citealt{1995MNRAS.274..602Q} for details). In the process we
assumed an exponential law for the radial velocity dispersion:
\begin{equation} \label{eq:sig}
  \sigma_{R} =\sigma_{0} +  \sigma_{1} e^{-\frac{R}{R_{\sigma}}},
\end{equation}
Our models, therefore, have three free parameters, $\sigma_{0}$,
$\sigma_{1}$, and $R_{\sigma}$, and varying them we constructed gaseous
disc models of NGC~2974. The models were compared with the symmetrised
velocity and velocity dispersion maps averaging the four symmetric
positions on the maps as required for an axisymmetric map
(Section~\ref{ss:model}). The best-fitting model was obtained for
$\sigma_{0} = 85 \pm 5\kms$, $\sigma_{1} = 180 \pm 10\kms$, and
$R_{\sigma} = 5 \pm 1\arcsec$. Comparison of this model with the
symmetrised observations is presented in Fig. ~\ref{f:asym}.

This simple asymmetric drift model can reproduce rather well the
general properties of the emission-line gas disc, including the bulk
of the streaming velocity as well as the significant non-zero velocity
dispersion. The overall fit is quite good with a mean difference of
only 5\%, with peaks up to about 20\%.  The main discrepancy occurs
along the major-axis at a radius around 10\arcsec, close to where we
observe the presumed elliptical ring in the ionised gas equivalent
width (see Fig.~6).  The observed minor-axis elongation of the gas
velocity dispersion is also not reproduced by the model. This is not
surprising since some of those features, as mentioned, are signatures
of non-axisymmetry in NGC~2974, and cannot be represented by simple
axisymmetric models.

Since the normalisation of the kinematics is mostly controlled by the
assumed mass-to-light ratio, the goodness of the fit implies that the
overall energy budget coming from the Schwarzschild modelling is
consistent with the observed gas kinematics: the high value of the gas
velocity dispersion is therefore globally compensated by a lower mean
velocity.  A similar finding was reported by
\citet{1994MNRAS.270..325C}.  All this therefore raises the question
of how the gas can remain dynamically hot in a disc galaxy like
NGC~2974 at a scale of $\sim 1$~kpc.

%
%

\section{Concluding remarks}
\label{s:con}
This paper presents a case study of the early-type galaxy NGC~2974,
which was observed in the course of the \SAURON survey of nearby E/S0
galaxies.

Kinematic position angles of the stellar and gaseous kinematics of
this galaxy are on average well aligned. The stellar kinematic maps
exhibit point- and mirror-(anti)-symmetry, with the kinematic angle
equal to the photometric PA, and are consistent with an axisymmetric
intrinsic shape. The gaseous velocity map is more complicated, with
clear departures from axisymmetry in the centre of the galaxy
($<4\arcsec$). At larger radii, the gas kinematic angle is not
constant, although is largely consistent with the photometric PA,
showing deviations of a few degrees. The \SAURON observations of
NGC~2974 confirm the existence of non-axisymmetric perturbations
consistent with an inner bar (EGF03\nocite{2003MNRAS.345.1297E}) as
well as a possible large-scale bar. The departures from axisymmetry
are not visible in the stellar kinematic maps and therefore are likely
to be weak. This allows us to construct axisymmetric models of
NGC~2974.

We constructed self-consistent three-integral axisymmetric models
based on the Schwarzschild's orbit superposition method, varying
mass-to-light ratio and inclination. The observed surface brightness
was parameterised by multi-Gaussian expansion model on both ground-
and space-based imaging. The models were compared with the \SAURON
kinematic maps of the first six moments of the LOSVD ($v$, $\sigma$,
$h_{3}$ - $h_{6}$). The best-fitting model has $\Upsilon=4.5 \pm 0.1$
and $i=65\degr \pm 2.5\degr$. The inclination is formally well
constrained, but there are several indications that the recovery of
the inclination is uncertain: (i) differences between the models are
on the level of the expected systematics in the data (e.g. template
mismatch); (ii) difference between the best-fit model and the data are
bigger than differences between other models and the best-fit model;
(iii) limitations of the models (discreteness of the cusps) may
artificially constrain the inclination.

The internal structure of NGC~2974 (assuming axisymmetry) reveals the
existence of a rapidly rotating component contributing with about 10\%
of the total light. This component is composed of orbits allowing the
third integral and does not represent a cold stellar disc, although
suggests a flattened structure similar to an S0 galaxy.

The results of the stellar dynamical models were compared with the
results of modelling the gas component. The inclination of the gas
disc, calculated from the emission-line velocity map is $i=60\degr \pm
3$, in agreement with the formally constrained stellar inclination. A
simple model of the gas disc in the same potential used for the
stellar modelling ($\Upsilon=4.5$, but $i=60\degr$) was able to
reproduce the characteristics of the gas kinematics ($v$ and $\sigma$)
on the large scale, but, as expected, failed in the centre. In
general, the observed gas kinematics is consistent with being produced
by gravitational potential determined by the stellar dynamical models,
but the origin of the high velocity dispersion of the gas on large
scales remains an open question.

We performed a set of tests of our implementation of the
Schwarzschild's orbit superposition method. For this purpose we
constructed a general two-integral model of NGC~2974, and used the
reconstructed kinematics as inputs to the Schwarzschild's method. We
tested (i) the influence of the radial coverage of the kinematic data
on the internal structure, (ii) the recovery of the test model
parameters ($\Upsilon$,$i$), and (iii) the recovery of the test model
DF. The tests show that:

\begin{enumerate}
\item Increasing the radial coverage of the kinematic data from $1r_e$
      to $2r_e$ does not change the internal structure within
      $1r_e$. The results of the dynamical models of the \SAURON
      observations of NGC~2974 would not change if the radial coverage
      would be increased by a factor of 2.

\item We find that three-integral models can accurately recover the
      mass-to-light ratio. Although the models are also able to
      constrain the inclination of the test model formally, the
      apparent differences between the models are small (as in the
      case of the real observations). Under careful examination, it is
      possible to choose the best model by eye, but the decisive
      kinematic features are below (or at the level) of the expected
      systematics in the data (e.g. template mismatch) and might be
      influenced by the uncertainties in the models
      (e.g. regularisation or variations in the sampling of
      observables with orbits). This suggest a degeneracy of models
      with respect to the recovery of inclination. More general tests
      on other galaxies and theoretical work is needed for a better
      understanding of this issue.

\item From a realistic test model, the analytically known input DF is
      recovered within 6\% in the region constrained
      by integral-field kinematics. This suggests that applying the
      Schwarzschild technique with integral-field kinematics can
      reliably recover a representative DF.

\end{enumerate}

\vspace{+1cm}
\noindent{\bf Acknowledgements}\\

We thank Glenn van de Ven for fruitful discussions about the recovery
of the DF. DK was supported by NOVA, the Netherlands Research school
for Astronomy. MC acknowledges support from a VENI grant award by the
Netherlands Organization of Scientific Research (NWO).

\label{lastpage}

\begin{thebibliography}{}

\bibitem[\protect\citeauthoryear{{Aguerri}, {Debattista} \&
  {Corsini}}{{Aguerri} et~al.}{2003}]{2003MNRAS.338..465A}
{Aguerri} J.~A.~L.,  {Debattista} V.~P.,    {Corsini} E.~M.,  2003, \mnras,
  338, 465

\bibitem[\protect\citeauthoryear{{Amico}, {Bertin}, {Bertola},
  {Buson}, {Danziger}, {Dejonghe}, {Pizzella}, {Sadler}, {Saglia},
  {Stiavelli}, {de Zeeuw} \& {Zeilinger}}{{Amico}
  et~al.}{1993}]{1993sdce.conf..225A}
{Amico} P., et~al., 1993, in Danziger I,~J., Zellinger W.~W., Kjaerp
  K., eds, Structure, Dynamics and Chemical Evolution of Elliptical
  Galaxies. ESO, Garching, p. 225

\bibitem[\protect\citeauthoryear{{Bacon}, {Copin}, {Monnet}, {Miller},
  {Allington-Smith}, {Bureau}, {Carollo}, {Davies},
  {Emsellem}, {Kuntschner}, {Peletier}, {Verolme} \& {Tim de
  Zeeuw}}{{Bacon} et~al.}{2001}]{2001MNRAS.326...23B}
{Bacon} R., et~al., 2001, \mnras, 326, 23

\bibitem[\protect\citeauthoryear{{Barth}, {Sarzi}, {Rix}, {Ho}, {Filippenko} \&
  {Sargent}}{{Barth} et~al.}{2001}]{2001ApJ...555..685B}
{Barth} A.~J.,  {Sarzi} M.,  {Rix} H.,  {Ho} L.~C.,  {Filippenko} A.~V.,
  {Sargent} W.~L.~W.,  2001, \apj, 555, 685

\bibitem[\protect\citeauthoryear{{Bender}}{{Bender}}{1988}]{1988A&A...193L...7%
B}
{Bender} R.,  1988, \aap, 193, L7

\bibitem[\protect\citeauthoryear{{Bertola}, {Cinzano}, {Corsini}, {Rix} \&
  {Zeilinger}}{{Bertola} et~al.}{1995}]{1995ApJ...448L..13B}
{Bertola} F.,  {Cinzano} P.,  {Corsini} E.~M.,  {Rix} H.,    {Zeilinger} W.~W.,
   1995, \apjl, 448, L13+

\bibitem[\protect\citeauthoryear{{Binney} \& {Tremaine}}{{Binney} \&
  {Tremaine}}{1987}]{1987gady.book.....B}
{Binney} J.,  {Tremaine} S.,  1987, {Galactic dynamics}.
Princeton, NJ, Princeton University Press, 1987, 747 p.

\bibitem[\protect\citeauthoryear{{Bregman}, {Hogg} \& {Roberts}}{{Bregman}
  et~al.}{1992}]{1992ApJ...387..484B}
{Bregman} J.~N.,  {Hogg} D.~E.,    {Roberts} M.~S.,  1992, \apj, 387, 484

\bibitem[\protect\citeauthoryear{{Buson}, {Sadler}, {Zeilinger}, {Bertin},
  {Bertola}, {Danzinger}, {Dejonghe}, {Saglia} \& {de Zeeuw}}{{Buson}
  et~al.}{1993}]{1993A&A...280..409B}
{Buson} L.~M., et~al., 1993, \aap, 280, 409

\bibitem[\protect\citeauthoryear{{Cappellari}}{{Cappellari}}{2002}]{2002MNRAS.%
333..400C}
{Cappellari} M.,  2002, \mnras, 333, 400

\bibitem[\protect\citeauthoryear{{Cappellari} \& {Copin}}{{Cappellari} \&
  {Copin}}{2003}]{2003MNRAS.342..345C}
{Cappellari} M.,  {Copin} Y.,  2003, \mnras, 342, 345

\bibitem[\protect\citeauthoryear{{Cappellari} \& {Emsellem}}{{Cappellari} \&
  {Emsellem}}{2004}]{2004PASP..116..138C}
{Cappellari} M.,  {Emsellem} E.,  2004, \pasp, 116, 138

\bibitem[\protect\citeauthoryear{{Cappellari}, {Verolme}, {van der Marel},
  {Kleijn}, {Illingworth}, {Franx}, {Carollo} \& {de Zeeuw}}{{Cappellari}
  et~al.}{2002}]{2002ApJ...578..787C}
{Cappellari} M.,  {Verolme} E.~K.,  {van der Marel} R.~P.,  {Kleijn} G.~A.~V.,
  {Illingworth} G.~D.,  {Franx} M.,  {Carollo} C.~M.,    {de Zeeuw} P.~T.,
  2002, \apj, 578, 787

\bibitem[Cappellari et~al.(2004)]{2004cbhg.sympE...5C} Cappellari M.,
et al., 2004, in Ho L. C., ed, Carnegie Observatories Astrophysics
Series, Vol. 1: Coevolution of Black Holes and Galaxies, (Pasadena:
Carnegie Observatories;
http://www.ociw.edu/ociw/symposia/series/\\symposium1/proceedings.html)

\bibitem[\protect\citeauthoryear{{Cinzano} \& {van der Marel}}{{Cinzano} \&
  {van der Marel}}{1994}]{1994MNRAS.270..325C}
{Cinzano} P.,  {van der Marel} R.~P.,  1994, \mnras, 270, 325

\bibitem[\protect\citeauthoryear{{Copin}, {Bacon}, {Bureau}, {Davies},
  {Emsellem}, {Kuntschner}, {Miller}, {Peletier}, {Verolme} \& {de
  Zeeuw}}{{Copin} et~al.}{2001}]{2001sf2a.conf..289C}
{Copin} Y., et~al., 2001, in SF2A-2001: Semaine de l'Astrophysique
  Francaise {Kinemetry: quantifying kinematic maps}.  p. 289

\bibitem[\protect\citeauthoryear{{Copin}, {Cretton} \& {Emsellem}}{{Copin}
  et~al.}{2004}]{2004A&A...415..889C}
{Copin} Y.,  {Cretton} N.,    {Emsellem} E.,  2004, \aap, 415, 889

\bibitem[\protect\citeauthoryear{{Cretton}, {de Zeeuw}, {van der Marel} \&
  {Rix}}{{Cretton} et~al.}{1999}]{1999ApJS..124..383C}
{Cretton} N.,  {de Zeeuw} P.~T.,  {van der Marel} R.~P.,    {Rix} H.,  1999,
  \apjs, 124, 383

\bibitem[\protect\citeauthoryear{{Cretton} \& {Emsellem}}{{Cretton} \&
  {Emsellem}}{2004}]{2004MNRAS.347L..31C}
{Cretton} N.,  {Emsellem} E.,  2004, \mnras, 347, L31

\bibitem[\protect\citeauthoryear{{Cretton}, {Rix} \& {de Zeeuw}}{{Cretton}
  et~al.}{2000}]{2000ApJ...536..319C}
{Cretton} N.,  {Rix} H.,    {de Zeeuw} P.~T.,  2000, \apj, 536, 319

\bibitem[\protect\citeauthoryear{{Cretton} \& {van den Bosch}}{{Cretton} \&
  {van den Bosch}}{1999}]{1999ApJ...514..704C}
{Cretton} N.,  {van den Bosch} F.~C.,  1999, \apj, 514, 704

\bibitem[\protect\citeauthoryear{{de Zeeuw}, {Bureau}, {Emsellem}, {Bacon},
  {Marcella Carollo}, {Copin}, {Davies}, {Kuntschner}, {Miller}, {Monnet},
  {Peletier} \& {Verolme}}{{de Zeeuw} et~al.}{2002}]{2002MNRAS.329..513D}
{de Zeeuw} P.~T., et~al., 2002, \mnras, 329, 513

\bibitem[\protect\citeauthoryear{{Debattista} \& {Williams}}{{Debattista} \&
  {Williams}}{2004}]{2004ApJ...605..714D}
{Debattista} V.~P.,  {Williams} T.~B.,  2004, \apj, 605, 714

\bibitem[\protect\citeauthoryear{{Emsellem}, {Cappellari}, {Peletier},
  {McDermid}, {Geacon}, {Bureau}, {Copin}, {Davies}, {Krajnovi{\' c}},
  {Kuntschner}, {Miller} \& {Tim de Zeeuw}}{{Emsellem}
  et~al.}{2004}]{2004MNRAS.352..721E}
{Emsellem} E., et~al., 2004, \mnras, 352, 721

\bibitem[\protect\citeauthoryear{{Emsellem}, {Dejonghe} \& {Bacon}}{{Emsellem}
  et~al.}{1999}]{1999MNRAS.303..495E}
{Emsellem} E.,  {Dejonghe} H.,    {Bacon} R.,  1999, \mnras, 303, 495

\bibitem[\protect\citeauthoryear{{Emsellem}, {Goudfrooij} \&
  {Ferruit}}{{Emsellem} et~al.}{2003}]{2003MNRAS.345.1297E}
{Emsellem} E.,  {Goudfrooij} P.,    {Ferruit} P.,  2003, \mnras, 345, 1297

\bibitem[\protect\citeauthoryear{{Emsellem}, {Monnet} \& {Bacon}}{{Emsellem}
  et~al.}{1994}]{1994A&A...285..723E}
{Emsellem} E.,  {Monnet} G.,    {Bacon} R.,  1994, \aap, 285, 723

\bibitem[\protect\citeauthoryear{{Erwin} \& {Sparke}}{{Erwin} \&
  {Sparke}}{2002}]{2002AJ....124...65E}
{Erwin} P.,  {Sparke} L.~S.,  2002, \aj, 124, 65

\bibitem[\protect\citeauthoryear{{Erwin} \& {Sparke}}{{Erwin} \&
  {Sparke}}{2003}]{2003ApJS..146..299E}
{Erwin} P.,  {Sparke} L.~S.,  2003, \apjs, 146, 299

\bibitem[\protect\citeauthoryear{{Franx}, {van Gorkom} \& {de Zeeuw}}{{Franx}
  et~al.}{1994}]{1994ApJ...436..642F}
{Franx} M.,  {van Gorkom} J.~H.,    {de Zeeuw} T.,  1994, \apj, 436, 642

\bibitem[\protect\citeauthoryear{{Friedli} \& {Martinet}}{{Friedli} \&
  {Martinet}}{1993}]{1993A&A...277...27F}
{Friedli} D.,  {Martinet} L.,  1993, \aap, 277, 27

\bibitem[\protect\citeauthoryear{{Gebhardt}, {Richstone}, {Tremaine}, {Lauer},
  {Bender}, {Bower}, {Dressler}, {Faber}, {Filippenko}, {Green}, {Grillmair},
  {Ho}, {Kormendy}, {Magorrian} \& {Pinkney}}{{Gebhardt}
  et~al.}{2003}]{2003ApJ...583...92G}
{Gebhardt} K., et~al., 2003, \apj, 583, 92

\bibitem[\protect\citeauthoryear{{Gerhard}}{{Gerhard}}{1993}]{1993MNRAS.265..2%
13G}
{Gerhard} O.~E.,  1993, \mnras, 265, 213

\bibitem[\protect\citeauthoryear{{Goudfrooij}, {Hansen}, {Jorgensen},
  {Norgaard-Nielsen}, {de Jong} \& {van den Hoek}}{{Goudfrooij}
  et~al.}{1994}]{1994A&AS..104..179G}
{Goudfrooij} P.,  {Hansen} L.,  {Jorgensen} H.~E.,  {Norgaard-Nielsen} H.~U.,
  {de Jong} T.,    {van den Hoek} L.~B.,  1994, \aaps, 104, 179

\bibitem[\protect\citeauthoryear{{Hunter} \& {Qian}}{{Hunter} \&
  {Qian}}{1993}]{1993MNRAS.262..401H}
{Hunter} C.,  {Qian} E.,  1993, \mnras, 262, 401

\bibitem[\protect\citeauthoryear{{Kim}}{{Kim}}{1989}]{1989ApJ...346..653K}
{Kim} D.,  1989, \apj, 346, 653

\bibitem[\protect\citeauthoryear{{Kim}, {Jura}, {Guhathakurta}, {Knapp} \& {van
  Gorkom}}{{Kim} et~al.}{1988}]{1988ApJ...330..684K}
{Kim} D.-W.,  {Jura} M.,  {Guhathakurta} P.,  {Knapp} G.~R.,    {van Gorkom}
  J.~H.,  1988, \apj, 330, 684

\bibitem[\protect\citeauthoryear{{Laine}, {Shlosman}, {Knapen} \&
       {Peletier}}{{Laine} et~al.}{2002}]{2002ApJ...567...97L} Laine,
       S., Shlosman, I., Knapen, J.~H., \& Peletier, R.~F.\ 2002, \apj,
       567, 97

\bibitem[\protect\citeauthoryear{{Lawson} \& {Hanson}}{{Lawson} \&
  {Hanson}}{1974}]{1974slsp.book.....L}
{Lawson} C.~L.,  {Hanson} R.~J.,  1974, {Solving least squares problems}.
Prentice-Hall Series in Automatic Computation, Prentice-Hall, Englewood Cliffs

\bibitem[\protect\citeauthoryear{{Osterbrock}}{{Osterbrock}}{1989}]{1989agna.b%
ook.....O}
{Osterbrock} D.~E., 1989, {Astrophysics of gaseous nebulae and active
  galactic nuclei}., University Science Books, Sausalito, CA

\bibitem[\protect\citeauthoryear{{Pfenniger} \& {Norman}}{{Pfenniger} \&
  {Norman}}{1990}]{1990ApJ...363..391P}
{Pfenniger} D.,  {Norman} C.,  1990, \apj, 363, 391

\bibitem[\protect\citeauthoryear{{Plana}, {Boulesteix}, {Amram}, {Carignan} \&
  {Mendes de Oliveira}}{{Plana} et~al.}{1998}]{1998A&AS..128...75P}
{Plana} H.,  {Boulesteix} J.,  {Amram} P.,  {Carignan} C.,    {Mendes de
  Oliveira} C.,  1998, \aaps, 128, 75

\bibitem[Press et al.(1992)]{pre92} Press W.~H., Teukolsky S.~A.,
Vetterling W.~T., \& Flannery B.~P., 1992, Numerical Recipes in
FORTRAN 77. 2d ed, Cambridge Univ. Press, Cambridge

\bibitem[\protect\citeauthoryear{{Qian}, {de Zeeuw}, {van der Marel} \&
  {Hunter}}{{Qian} et~al.}{1995}]{1995MNRAS.274..602Q}
{Qian} E.~E.,  {de Zeeuw} P.~T.,  {van der Marel} R.~P.,    {Hunter} C.,  1995,
  \mnras, 274, 602

\bibitem[\protect\citeauthoryear{{Rix}, {de Zeeuw}, {Cretton}, {van der Marel}
  \& {Carollo}}{{Rix} et~al.}{1997}]{1997ApJ...488..702R}
{Rix} H.,  {de Zeeuw} P.~T.,  {Cretton} N.,  {van der Marel} R.~P.,
  {Carollo} C.~M.,  1997, \apj, 488, 702

\bibitem[\protect\citeauthoryear{{Rix} \& {White}}{{Rix} \&
  {White}}{1992}]{1992MNRAS.254..389R}
{Rix} H.,  {White} S.~D.~M.,  1992, \mnras, 254, 389

\bibitem[\protect\citeauthoryear{{Rybicki}}{{Rybicki}}{1987}]{1987IAUS..127..3%
97R}
{Rybicki} G.~B., 1987, in de Zeeuw P.~T., ed, IAU Symp. 127:
  Structure and Dynamics of Elliptical Galaxies, p 397, Reidel,
  Dordrecht

\bibitem[\protect\citeauthoryear{{Schoenmakers}, {Franx} \& {de
  Zeeuw}}{{Schoenmakers} et~al.}{1997}]{1997MNRAS.292..349S}
{Schoenmakers} R.~H.~M.,  {Franx} M.,    {de Zeeuw} P.~T.,  1997, \mnras, 292,
  349

\bibitem[\protect\citeauthoryear{{Schwarzschild}}{{Schwarzschild}}{1979}]{1979%
ApJ...232..236S}
{Schwarzschild} M.,  1979, \apj, 232, 236

\bibitem[\protect\citeauthoryear{{Schwarzschild}}{{Schwarzschild}}{1982}]{1982%
ApJ...263..599S}
{Schwarzschild} M.,  1982, \apj, 263, 599

\bibitem[\protect\citeauthoryear{{Thomas}, {Saglia}, {Bender},
{Thomas}, D. {Gebhardt}, {Magorrian} \& {Richstone}}{{Thomas}
et~al.}{2004}]{2004MNRAS.353..391T} {Thomas}, J., {Saglia}, R.~P.,
{Bender}, R., {Thomas}, D., {Gebhardt}, K., {Magorrian}, J.,
{Richstone}, D., 2004, \mnras, 353, 391

\bibitem[\protect\citeauthoryear{{Tonry}, {Dressler}, {Blakeslee}, {Ajhar},
  {Fletcher}, {Luppino}, {Metzger} \& {Moore}}{{Tonry}
  et~al.}{2001}]{2001ApJ...546..681T}
{Tonry} J.~L.,  {Dressler} A.,  {Blakeslee} J.~P.,  {Ajhar} E.~A.,  {Fletcher}
  A.~B.,  {Luppino} G.~A.,  {Metzger} M.~R.,    {Moore} C.~B.,  2001, \apj,
  546, 681

\bibitem[\protect\citeauthoryear{{Tremaine}, {Gebhardt}, {Bender}, {Bower},
  {Dressler}, {Faber}, {Filippenko}, {Green}, {Grillmair}, {Ho}, {Kormendy},
  {Lauer}, {Magorrian}, {Pinkney} \& {Richstone}}{{Tremaine}
  et~al.}{2002}]{2002ApJ...574..740T}
{Tremaine} S., et~al., 2002, \apj, 574, 740

\bibitem[\protect\citeauthoryear{{Valluri}, {Merritt} \& {Emsellem}}{{Valluri}
  et~al.}{2004}]{2004ApJ...602...66V}
{Valluri} M.,  {Merritt} D.,    {Emsellem} E.,  2004, \apj, 602, 66

\bibitem[\protect\citeauthoryear{{van der Marel}, {Cretton}, {de Zeeuw} \&
  {Rix}}{{van der Marel} et~al.}{1998}]{1998ApJ...493..613V}
{van der Marel} R.~P.,  {Cretton} N.,  {de Zeeuw} P.~T.,    {Rix} H.,  1998,
  \apj, 493, 613

\bibitem[\protect\citeauthoryear{{van der Marel} \& {Franx}}{{van der Marel} \&
  {Franx}}{1993}]{1993ApJ...407..525V}
{van der Marel} R.~P.,  {Franx} M.,  1993, \apj, 407, 525

\bibitem[\protect\citeauthoryear{{van der Marel} \& {van den Bosch}}{{van der
  Marel} \& {van den Bosch}}{1998}]{1998AJ....116.2220V}
{van der Marel} R.~P.,  {van den Bosch} F.~C.,  1998, \aj, 116, 2220

\bibitem[\protect\citeauthoryear{{Vandervoort}}{{Vandervoort}}{1984}]{1984ApJ.%
..287..475V}
{Vandervoort} P.~O.,  1984, \apj, 287, 475

\bibitem[\protect\citeauthoryear{{Verdoes Kleijn}, {van der Marel}, {Carollo}
  \& {de Zeeuw}}{{Verdoes Kleijn} et~al.}{2000}]{2000AJ....120.1221V}
{Verdoes Kleijn} G.~A.,  {van der Marel} R.~P.,  {Carollo} C.~M.,    {de Zeeuw}
  P.~T.,  2000, \aj, 120, 1221

\bibitem[\protect\citeauthoryear{Verolme \& de Zeeuw}{2002}]{ver02}
Verolme E.~K., de Zeeuw P.~T., 2002, \mnras, 331, 959

\bibitem[\protect\citeauthoryear{{Verolme}, {Cappellari}, {Copin},
  {van der Marel}, {Bacon}, {Bureau}, {Davies}, {Miller} \& {de
  Zeeuw}}{{Verolme} et~al.}{2002}]{2002MNRAS.335..517V}
{Verolme} E.~K., et~al., 2002, \mnras, 335, 517

\bibitem[\protect\citeauthoryear{{Wong}, {Blitz} \& {Bosma}}{{Wong}
  et~al.}{2004}]{2004ApJ...605..183W}
{Wong} T.,  {Blitz} L.,    {Bosma} A.,  2004, \apj, 605, 183

\end{thebibliography}
\end{document}